# Mitigating Imperfections in Mixed-Signal Neuromorphic Circuits

Z. Fahimi[†*], M. R. Mahmoodi[†*], M. Klachko[†], H. Nili, H. Kim, and D. B. Strukov
Department of Electrical and Computer Engineering, University of California, Santa Barbara, CA 93106, USA
*{zfahimi, mrmahmoodi}@ucsb.edu

The progress in neuromorphic computing is fueled by the development of novel nonvolatile memories capable of storing analog information and implementing neural computation efficiently. However, like most other analog circuits, these devices and circuits are prone to imperfections, such as temperature dependency, noise, tuning error, etc., often leading to considerable performance degradation in neural network implementations. Indeed, imperfections are major obstacles in the path of further progress and ultimate commercialization of these technologies. Hence, a practically viable approach should be developed to deal with these nonidealities and unleash the full potential of nonvolatile memories in neuromorphic systems. Here, for the first time, we report a comprehensive characterization of critical imperfections in two analog-grade memories, namely passively-integrated memristors and redesigned eFlash memories, which both feature long-term retention, high endurance, analog storage, low-power operation, and compact nano-scale footprint. Then, we propose a holistic approach that includes modifications in the training, tuning algorithm, memory state optimization, and circuit design to mitigate these imperfections. Our proposed methodology is corroborated on a hybrid software/experimental framework using two benchmarks: a moderate-size convolutional neural network and ResNet-18 trained on CIFAR-10 and ImageNet datasets, respectively. Our proposed approaches allow 2.5× to 9× improvements in the energy consumption of memory arrays during inference and sub-percent accuracy drop across 25–100 °C temperature range. The defect tolerance is improved by >100×, and a sub-percent accuracy drop is demonstrated in deep neural networks built with 64×64 passive memristive crossbars featuring 25% normalized switching threshold variations. We believe that our results substantially improve the inference accuracy and efficiency of neuromorphic systems and paves the way towards building reliable large-scale integrated neuromorphic circuits.

## Introduction

A tremendous body of research predicts analog implementations of neuromorphic networks [1,4,5,6] could bridge the gap between artificial and biological prototypes [7,8,9] and offer comparable areal density to biological prototypes at a better processing time. The background for this improvement is the physical level implementation of the vector-by-matrix multiplication (VMM), the frequent operation in neuromorphic networks, and the efficient realization of an analog synapse capable of both holding a learnable parameter (or weight) and performing useful computations (Fig. 1a). Inventing a reliable synaptic device has been the main obstacle in achieving analog computing supremacy. The recent development of continuous-state nonvolatile memory synapses is perhaps a milestone that paves the way for achieving this goal [10]. The most notable candidates that excel in primary features such as long-term retention, high endurance, analog storage, low-power operation, and compact footprint are metal-oxide passive memristors [11] (Fig. 1b) and redesigned eFlash memories (Fig. 1c) [12]. Nevertheless, all synaptic devices are generally more or less prone to imperfection such as temperature dependency, noise, yield, drift, tuning error, and static nonlinearity. While imperfections are not necessarily meant to be detrimental (see, e.g., [13]), they severely degrade the accuracy of currently popular DNNs.

The endeavors to improve device reliability are ongoing and actively pursued. A massive number of works focus on improving synapse reliability by harnessing novel materials and stacks, e.g., reducing noise [14], enhancing uniformity [15,16], linearity [17], which is vividly the most promising approach in the long run. In the meantime, a large body of research explores circuit, system, and algorithmic techniques to mitigate these nonidealities. Such efforts are categorized into 5 approaches, as shown in Fig. 1.

[†] These authors contributed equally to this work.

The most predominant approach in implementing neuromorphic systems is through ex-situ training, in which synaptic weights are calculated on a precursor server [18]. The synaptic weights are then transferred to numerous mixed-signal chips, which only support the inference task. A trivial approach to cope with imperfections is by adding redundancy [19]: the model and algorithm hyperparameters are selected such that the deployed model in the analog domain would tolerate a certain amount of unreliability. For example, an enormous network such as AlexNet can endure a large amount of noise that allows weight binarization [20]. In more compact models, it might be possible to lower computing precision without any impact on accuracy at the cost of redundancy. For example, the accuracy loss by changing weight precision from 4-bit to 2-bit in ResNet-18 can be compensated by doubling the model size [21]. Ref. [22] proposes to enlarge the capacity of a fully connected network through committee machines and validates the results on the MNIST dataset. However, the major concern with this approach is the lack of evidence that this approach is scalable to complicated tasks.

The second approach in dealing with imperfections is through in-situ training [23,4], in which the training and inference are both performed on the mixed-signal hardware. The first drawback is the substantial areal overhead needed for the infrequent training operations, e.g., to compute and store gradients. Besides, while the chip could become resilient to some imperfections, other nonidealities associated with the dynamic behavior of devices may arise. Some works propose a hybrid approach that imposes less resource overhead, e.g., the model is initially trained ex-situ, and then an online calibration scheme modifies the weights in the run time. For example, Ref. [24] proposes an adaptive batch normalization technique that effectively compensates for the retention loss in memory cells. The drift in phase-change memories is high (~50% conductance drift after 100 ms [25]) that an always-on compensation circuit is required. Another solution is through chip-in-the-loop ex-situ training [26-28], in which specific features are measured for an individual chip (e.g., faulty synapse locations [29] or drift statistics [30]) and then applied in the training phase in the server. The adapted weights are then transferred to the chip in the deployment phase. Chip-in-the-loop ex-situ training could also be implemented by running all forward pass operations in the target device and all backward pass operations in the GPU cluster. But strategies that include device avoidance/reconfiguration/remapping or are chip-specific might present scalability challenges, despite the ability to boost the performance of an individual chip.

On the other hand, hardware-aware ex-situ training is a more scalable method in which hardware nonidealities are modeled and included in the training phase to generate a robust model. Ref. [18] uses device noise models during the training to improve the robustness of a moderate-size mixed-signal convolutional network. In Ref. [31], DropConnect regularization is introduced to enhance the accuracy drop originating from low yield without using nonideality models—at 98% yield, 17% accuracy drop on CIFAR-10 based on ResNet-18 model is reduced to 10%. The majority of other previous works either study a particular nonideality [29,32-33] or consider redundant networks on smaller datasets [35,36]. Besides, some focus solely on simulations with no experimental data or employ practically nonviable devices for modeling purposes.

In this paper, major imperfections on two prospective analog-grade synaptic devices, passively-integrated memristors, and eFlash memories, are characterized in order to determine nonidealities that severely impact mixed-signal DNNs. The choice of these promising technologies stems from the fact that, unlike most emerging technologies, they transcend all rudimentary features, including high endurance, analog storage, long-term retention, low-power operation, and nano-scale footprint (see supplementary Table 1 in [37] for a comparison). Besides, low conductance and switching voltage range and very dense cell size ($4F^2$ and ~110 $F^2$ for memristors and eFlash) allow practical implementation of large-scale mixed-signal DNNs with decently large VMMs (e.g., 64×64) [38].

In the case of eFlash memories, high precision tuning and superb analog-grade retention are reported, and excellent yield is deemed due to the maturity of the technology, making temperature variations and noise the major issues. Passive $TiO_2$ memristive technology also offers high analog retention in an excellent areal density despite susceptibility to temperature variations, noise, limited yield, and half select disturbance. Each factor is studied separately, and a holistic approach is proposed that includes modifications in the training, tuning, state optimization, and circuits and targets each issue individually. More importantly, our proposed method is practical in terms of implementation cost with negligible overhead and is validated on a hybrid experiment/simulation framework using two benchmarks: a moderate-size convolutional neural network (ConvNet) and ResNet-18 trained on CIFAR-10, and ImageNet datasets, respectively.

The accuracy drop is almost fully recovered in the 20 °C to 100 °C temperature range by employing three incrementally applied approaches: temperature-sweep batch training, *k*-reference batch normalization, and state optimization. Three techniques are also proposed to improve the accuracy in the presence of device noise given a fixed energy budget. A heuristic approach is developed to find an optimum signal-to-noise ratio (SNR) for each layer. A dynamic range optimization technique is proposed to adjust the output dynamic range of each layer, and finally, training with circuit noise is demonstrated to be effective. The models are also resilient against the minor static nonlinearity (dot-product nonlinearity, i.e., *IV* nonlinearity in memristors and subthreshold slope nonlinearity in eFlash). High precision individual-device tuning accuracy (<1%) is experimentally showed for both devices, but passive memristors suffer from half-select disturbance due to the lack of selector. We adopt a tuning algorithm to increase the tuning accuracy of memristive devices in the presence of large device-to-device variations. Besides, we show that the inclusion of tuning error distribution during the training process improves accuracy as well. Finally, two techniques are proposed to overcome the limited yield in emerging technologies, pair modification that minimizes the weight mapping error in the tuning phase and average error compensation that prevents the propagation of error through cascaded layers.

# Results

Two mainstream driving force technologies in neuromorphic circuits are emerging memristive crosspoint devices and industrial-grade redesigned eFlash memories [10]. The excellent density and scaling prospects of the former enable the efficient implementation of large DNNs. However, the slow advancing pace of this technology signifies immense fabrication challenges, e.g., high uniformity requirements in the *IV* characteristics of memristors. In our recent work [37], we report the successful development 64×64 passive crossbar circuit with record-breaking ~ 99% yield and <26% normalized uniformity and, for the first time, based on a foundry-compatible fabrication process. Evidenced by the promising results from the recent demonstrations of large-scale neural networks [39], the situation is much better for floating-gate devices due to the availability of industrial-grade eFlash embedded in most CMOS processes.

A comprehensive characterization of imperfections in both memory technologies is initially performed. The experimental measurements are then used to model the average behavior of the devices and circuits. A unified parameter to describe major nonidealities in both synapses is used: the relative error of the state current, $\Delta I/I_0$, where $I_0$ is the reference tuning current measured at the nominal biasing condition, and $\Delta I$ is the current deviation from the ideal behavior. The models are then incorporated into simulation platforms (PyTorch-based libraries) to predict the fidelity loss in the benchmarks. A comprehensive noise analysis that takes circuit topology, weight mapping, and device characteristics into account is also provided for studying the role of analog noise in the benchmarks.

## *Experimental Measurements*

Fig. 2a shows the scanning electron microscope image of the fabricated crossbar that includes 4096 $TiO_2$ memristors–see the Method section for a discussion on the fabrication process and relevant details on electroforming, tuning, and operation procedures. Fig. 2b shows the measured *IV* characteristics of 350 randomly selected devices in the non-disturbing low-voltage regime. Upon the application of a voltage in this regime (<0.5 V), the conductance (state) of crosspoint devices remains unchanged at a fixed voltage. However, due to the tunneling or thermionic emission charge transport mechanism, the devices become more conductive in higher voltages and hence nonlinear. Fig. 2c shows the average relative static nonlinearity error versus applied voltage for various conductance states. Fig. 2d shows the measurement results for the relative changes of conductance in 350 devices concerning variations in the die temperature (25–100 °C). The device conductance has proportional to absolute temperature and complementary to absolute temperature dependency in low conductive and high conductive states, respectively, due to the insulator-metal phase transition. In the case of our memristive devices, such transition occurs at ~70 µS, on average (Fig. 2d). We observe large errors, particularly in low conductive states, which could severely degrade the computational accuracy of mixed-signal models at elevated temperatures.

The switching characteristics of memristors determine how precisely we can adjust their conductances. Individually, we can tune a device with high accuracy, e.g., <1% relative error regardless of its initial conductance. The experimental results in Fig. 2e corroborate this observation on 50 randomly selected devices tuned to 1.7 µs, 50 µs, and 10 µs conductances consecutively. For each device, the accuracy is achieved in less than 100 pulses using a naïve write-verify algorithm. However, tuning dynamics are more complicated at the crossbar level since the half-select problem imposes disturbance on already tuned 0T1R memristors. Using additional gate lines in active crossbars (with dedicated in-cell selectors) solves this problem at the cost of at least two orders of magnitude increase in the cell size. Fig. 2f shows an example of the ultimate relative tuning error distribution after the entire 64×64 crossbar is programmed to the states that correspond to the grayscale quantized Einstein image [37]. The final tuning error distribution depends on the switching threshold distributions and the tuning algorithm.

To investigate the impact of long-term retention loss, we perform accelerated retention tests and use the Arrhenius equation for room temperature projection of the results. Fig. 2g shows the extremely stable analog-grade operation of 30 devices tuned in various states, subjected to 100 °C baking for >25 hours —translating into >14 years of room temperature operation assuming 1.1 eV activation energy [40]. Fig. 2h shows the distribution of relative retention loss error for 400 memristors after 14 years of projected room temperature operation. More details of the statistical analysis of data for different states are provided in Supplementary Figure 1. Interestingly, unlike binary memristors [47], the distribution of retention loss error is relatively symmetrical in midrange analog states, i.e., the devices could move toward higher or lower conductive states. Note that we also observe unidirectional retention loss in very high (shifting toward low conductive states) or low (shifting toward high conductive states) conductance states, but we generally avoid switching the devices to extreme values. Finally, Fig. 2i shows the corresponding standard deviation of the relative conductance change versus time binned to different states for these devices. The measured data show that the relative shift in conductance for most devices is expected to be <2% after several years of operation, which is adequately high for the practical implementation of ex-situ trained DNNs.

Fig. 3a shows the scanning electron microscope image of the fabricated redesigned eFlash memory array–see the Method section for a discussion on tuning and operation procedure. First, we measure the average static input/output characteristics of 200 synapses in the gate-coupled structure (peripheral devices are tuned to the maximum state current, $I_{\mathrm{max}}$=30 nA) and find the relative static nonlinearity error, which originates from the voltage-dependent capacitive coupling. Fig. 3b-c shows the static nonlinearity measurement results for multiple synaptic weights. The temperature dependency of state current is also measured and demonstrated in Fig. 3d for 100 eFlash cells tuned to various states. The corresponding relative weight error in the gate-couple structure is also provided in Fig. 3e, indicating significant errors in high temperatures, which could significantly impact the accuracy of neural circuits. The retention characteristics of 100 eFlash memories are measured at 100°C. The measurements are performed by tuning the devices to different states within the relevant dynamic range. Fig. 3f shows the stable operation of 25 devices at 100°C for >6 hours. Regardless of the initial state, we confirm that the relative state change for most devices is comparable with the noise floor of the measurement setup. This superior performance partially stems from much effort spent on optimizing the technology for industrial-grade applications. Finally, in Fig. 3h, the high precision tuning capability of eFlash memories is shown for 50 devices by tuning them with 1% targeted accuracy to 100 nA, 50 nA, 30 nA, and 15 nA, consecutively, each using less than 50 pulses.

The initial assessment of the experimental data indicates that the analog retention is promising in both devices; however, they are prone to variations in temperature that result in significant shifts in synaptic weights. Noise and static nonlinearity are fundamental bottlenecks in most analog systems, and neural circuits are no exception. In both eFlash- and memristor-based neuromorphic systems, we need to optimize the circuit with respect to noise and static nonlinearity. For redesigned eFlash cells, high precision tuning is obtained due to the redesigned memory cell [12], and excellent yield [39] is deemed due to the maturity of the technology. However, for passive memristors, the half-select disturbance bounds the weight tuning accuracy in neuromorphic circuits built with practically viable kernel sizes, and limited percent-scale yield is a major hindrance. These identified imperfections are then modeled to study their deleterious effect in massive neuromorphic networks simulated in the PyTorch environment.

### *Simulation Framework*

Supplementary Fig. 2 elaborates on the phenomenological modeling procedure for the temperature dependency of eFlash and memristors. Instead of using complex physics-based models that would significantly slow down the simulation time in our massive neuromorphic benchmarks, we use multi-order polynomial functions that efficiently predict the average behavior of devices. A sufficiently high fitting accuracy between experimental results and models is observed. Supplementary Fig. 3 shows high goodness-of-fit in modeling the static nonlinearity of both analog memory candidates and discusses how static nonlinearity varies with the tuning condition. Unlike previous works, we consider circuit topology, mapping, and device characteristics in the noise analysis and propose an input, weight, and topology-dependent methodology to simulate the noise in massive neuromorphic benchmarks. Supplementary Fig. 4 describes how these models are used to find the software-equivalent noise of the circuits. For memristive devices, the role of limited yield and half-select disturbance are studied as well. The former is performed by initially mapping every weight to the conductances of a pair of analog devices (differential structure) to perform signed computation. Then, given the yield probability, we use a uniform random number generator to pick random devices and alter their conductance to the minimum or maximum conductance range or a random state. These defect cases are observed in the experiments, stuck at low conductance happens when a device cannot be fully formed, stuck at high conductance occurs when a device cannot be reset right after the electroforming process, and stuck at random state turns up mainly due to the endurance failure. To study the impact of half-select disturbance, we need to emulate the conductance tuning algorithm and ex-situ weight transfer process in memristive VMMs. The switching behavior of 500 devices is measured in several initial conductances upon applying write voltage pulses with variable amplitudes. Supplementary Fig. 5 shows the modeling results and confirms that parameters closely reproduce the measurement results, and discusses how the model is used to emulate the tuning process.

## *Temperature Variations*

Temperature variations have the most drastic impact on mixed-signal neuromorphic circuits. The synaptic weights change dramatically with temperature, modulating the pre-activation signals of the neurons. Fig. 4a shows how the preactivations received by the first neuron in the fully-connected layer of ResNet-18 change with the temperature. The modulation of the pre-activation distributions occurs in all layers and neurons but with different rates. Fig. 4b shows the temperature dependency of multiple percentiles of the pre-activation distributions in 2 different layers. Interestingly, such shifts are almost monotonic in most neurons, partly because the conductance of synaptic devices (eFlash or memristors) changes monotonically with respect to the temperature (Supplementary Fig. 6 shows extended data for more layers and demonstrates various modulation rates).

Our proposed temperature compensation method consists of three incrementally applied approaches and aims to reduce the worst-case accuracy across the studied temperature range by modifying the circuit and training algorithm. The details of every technique are discussed in the Method section. The first approach is temperature-sweep batch training, in which we incorporate the temperature model of synapses in the training process. The training is performed such that the resultant model learns to deal with an average change of variations, and the worst-case error reduces moderately. Fig. 4c shows the reduction of the worst-case accuracy drop for different stacks and mappings in ResNet-18. For example, for RM1 (ReRAM stack, mapping 1), the worst-case drop (occurs at 100 °C) is reduced from ~66% to 23% after applying approach 1. The most optimum performance is also achieved in midrange temperatures (60 °C), as expected. The improvements in ConvNet are also encouraging since the worst-case drop is decreased from ~25% to 3.3% (for RM1), as shown in Fig. 4d. Inspired by our previous work on increasing the reliability of hardware security primitives [41], we adopt $k$-reference batch normalization that further enhances the performance by using $k$ temperature-optimized batch normalization parameters per neuron. Fig. 4c shows a considerable reduction of the worst-case drop in the ImageNet benchmarks after applying the second approach ($k = 4$), e.g., from ~23% to 1.25% for RM1. In the CIFAR-10 benchmark, the worst-case drop for RM1 cases decreases from ~3.3% to 1.22%, with only 3 reference points. As depicted in Fig. 4e (for ResNet-18) and Fig. 4f (for ConvNet), we can further improve the results by increasing the number of reference points; a sub-percent accuracy drop is achieved with a few references (depending on the stack and mapping). Note that the model is still trained ex-situ entirely with a negligible overhead ($2k$ parameters per neuron). Though the second approach significantly reduces the worst-case accuracy drop, if needed, we can improve the results even further by optimizing the weight mapping parameters ($I_{\min}, I_{\mathrm{b}}$) for each weight. Supplementary Fig. 7 numerically analyzes experimental data and shows a procedure for finding the quasi-optimum design parameters of each device stack and weight mapping functionality. For example, the quasi-optimum minimum synaptic current for a given weight is obtained by $I_{\min}(\mathrm{nA}) = \max(0, 3 - 3.75(|w|/w_{\max}))$ when using mapping 1 of eFlash memories with the dynamic range of 30 nA. The state optimization approach, combined with temperature-sweep batch training and $k$-reference batch normalization, recovers the accuracy drop significantly across the entire temperature range regardless of selected device or mapping. The worst-case accuracy drop in the full temperature range diminishes to ~0.4% in ResNet-18 ($k = 4$) and ~0.49% in ConvNet ($k = 3$) in RM1 case. Fig. 4e-f highlights that a sub-percent accuracy drop is easily feasible across the full temperature range in both benchmarks after applying the temperature compensation techniques.

## *Noise*

Noise is a fundamental bottleneck for achieving high accuracy in every analog computing system. Since we can trade the accuracy loss (due to the noise) with power, the key is to optimize the performance given an energy budget. To aid this analysis, we use an energy scaling factor (see

Supplementary Fig. 4 for more details), a unified parameter that allows us to conveniently analyze the circuit without delving into the details of changing bandwidth, power consumption, or dynamic range for every device or layer. The trade-off between accuracy and the energy scaling factor in ResNet-18 is shown in Fig. 5a (see Supplementary Fig. 8a for the trade-off on ConvNet). The higher the energy scaler factor, the higher the accuracy drop. Here, we propose three techniques to improve the accuracy, given a fixed energy budget. The details of these approaches are discussed in the Method section.

First, we observe that depending on the hyper-parameters and structure of a network, the signal-to-noise ratio (SNR) requirements for a small accuracy drop are different in every layer. Fig. 5b shows how the accuracy drop in ResNet-18 alters when a constant noise power is added only to a specific layer. Although some layers are computationally less intensive, they require a lower energy scaler. Hence, different energy budgets should be spent on various layers. Finding the most optimum parameters of all layers is tedious, especially in large models. This partially stems from the fact that each layer has a particular sensitivity to noise that alters when the assigned SNR in each layer changes. Fig. 5c shows this sensitivity for various layers of ResNet-18. A layer-wise SNR-optimization algorithm is proposed that assigns optimized energy scaling factors to each layer and simultaneously optimizes the accuracy and energy consumption of memory arrays. The core idea is to increment the assigned SNR in small layers that need higher precision and reduce it in large layers that require low precision while the total energy budget is kept constant.

We consider 4 simulation cases (each representing a fixed energy budget) to demonstrate the effectiveness of our method: C1, C2, and C3, correspond to -17%, -10%, and -5% accuracy drop, respectively, and C4 is the case with unity energy scaling factor in all layers. Fig. 5d compares the cumulative distribution of relative pre-activation error in an output neuron amongst different cases. The layer-wise SNR optimization approach narrows the error distribution and increases the accuracy by ~7%. Another dramatic improvement is feasible by optimizing the range of activation signals. Our analysis in Supplementary Fig. 4 shows how the output referred noise in each layer depends on the range of the input signal. Since postsynaptic neuron signal distributions have large outliers but otherwise fairly uniform distributions, the entire signal range is often underused. We use a progressive brute force search method (see the method section) to find the optimum range of activation signals for every layer. As shown in Fig. 5d, the relative error distribution for C1 narrows down after applying the second technique, and the accuracy improves further by 6.8%. The inevitable nonlinearity in neuromorphic circuits modulates the statistics of propagated noise and signal through the deep layers. Further improvement is enabled by retraining the network for several epochs with the semi-optimized energy scaling factors and signal ranges obtained from the first two techniques, included in the forward simulation pass, and fine-tuning the network parameters based on the expected noise statistics in the implemented physical hardware. For example, the last method reduces the accuracy drop for the C1 case further to 1.5%. In order to achieve a 1.5% accuracy drop without applying any of the proposed methods, the energy budget spent on synaptic devices should increase by 6.8×. More comprehensive results of using the proposed techniques on various cases and devices are provided in Fig. 5e (for ResNet-18) and Fig. 5f (for ConvNet). A considerable reduction of accuracy drop is observed equivalent to saving computing energy, ranging from 2.5× to 9× in ResNet-18 benchmark. In addition, Supplementary Fig. 8b-c shows how the average and standard deviation of relative error in the output neuron signals improve after applying each method.

### *Half-Select Disturbance*

Passive crossbars are conventionally tuned by programing the devices in sequential (raster scan) order, typically starting from the device located on the first column/row to the last one. Each device is tuned using the write-verify algorithm until the target programming accuracy is achieved. However, this naïve method leads to considerable weight mapping errors in large kernels after the entire crossbar is programmed (see Supplementary Fig. 5 for more information on the naïve tuning). The problem could be alleviated by rerunning the tuning procedure multiple rounds through the full crossbar. However, Supplementary Fig. 9a-c shows that when the normalized switching threshold variations are more than 15% or the crossbar size is larger than 32×32, the net half-select disturbance is large enough to create a large tail of disturbed devices, even after 10 rounds of retuning the kernels.

Our recently proposed tuning procedure [37], initially tested on a small-scale MLP, includes two effective techniques that are also adopted in this work. In the first approach, the write voltage amplitudes are limited to a certain voltage, which is decreased gradually within each tuning round. This technique results in better average tuning accuracy than the naïve method because of the gradual reduction of disturbance in every tuning round, though some very high threshold devices might become more deviated from their targets. In the second approach (in addition to the first technique), devices with high set (reset) switching thresholds are identified and switched to the highest (lowest) conductive state prior to executing the first tuning round. Then, we take advantage of the possibility to encode the same weight with different target conductances in the differential pair implementation. In every round, when tuning a disturbed device with a threshold higher than the maximum voltage limit imposed by the first approach, we adjust the state of the paired device rather than tuning the high voltage device. Application of the two steps reduces the tail of disturbed devices and dramatically improves the classification accuracy.

In Fig. 6, several case studies are simulated to display the performance trends. Specifically, we focus on practically viable design points, i.e., 15% to 40% normalized switching threshold variations and 64×64 and 128×128 crossbar sizes. The entire ex-situ training is simulated for 12 random instances of each design point. The improvement achieved with each approach is shown in Fig. 6a, which compares the cumulative distribution of the absolute relative tuning error among them. Note that the distribution is obtained after the tuning is over (i.e., at the end of the $10^{th}$ round) for the devices implementing the first layer of ResNet-18 and assuming 64×64 crossbars. For the demonstrated crossbar in this paper (25% normalized variations and 64×64 crossbar), the naïve method leads to ~8.9% (~18.5%) average accuracy drop in ConvNet (ResNet-18), which are improved to 0.4% (1.8%), using the second approach. To further enhance the performance, the weights are randomly disturbed prior to each update during the baseline training. Although this approach does not transform the tuning error distribution, it makes the network inherently resilient toward unpredictable perturbations of weights in the tuning process. The third approach fully recovers the 0.4% accuracy drop in ConvNet and diminishes it to sub percent for ResNet-18.

## *Defect Tolerance*

Two techniques are also adopted that increase the resiliency of the mixed-signal hardware against defective devices. Note that the information that a specific device is defective is only available during the tuning phase. Supplementary Fig. 10 shows how the accuracy drop increases with the surge of faulty devices. Specifically, when using mapping 1, the network becomes more sensitive to devices stuck at high conductance and less susceptible to stuck at low conductance. This stems from the fact that the weight distribution in these benchmarks is such that most devices are tuned near the reset state for mapping 1 and near the midrange state for mapping 2. Fig. 7 compares the results when considering all three fault cases happening with equal probabilities and shows that mapping 2 outperforms mapping 1 (since the error distribution is statistically smaller). In the first approach, we exploit the fact that each weight is mapped to a pair of memory devices, and regardless of the mapping function, we can retune either of the devices to minimize the mapping error (see the Method section for more details). Supplementary Fig. 10 shows the improvement achieved by this technique in every fault case individually, and Fig. 7 shows the result of the general case. In a process with $2\times10^4$ ppm defective devices, both ResNet-18 and ConvNet generate almost entirely random classes without applying this technique. The proposed method diminishes the accuracy drop to only 14.3% for ConvNet and 23.4% for ResNet-18.

When a synapse is defective, it can potentially create a large shift in the average of preactivations in part due to the ununiform distribution of input activations, and more importantly, the limited dynamic range of the preactivations in properly trained networks. Further improvements could be achieved by compensating for such shifts. In every kernel, a pair of extra memory devices are included per neuron, which are tuned to remove the average error induced by the faulty devices (see the Method section for more details). Note that the area overhead of this method is negligible (unlike previous attempts to overcome this issue by adding redundancy), as there is no need for an additional or general-purpose routing at the input or output of the kernels. A fixed input always drives the extra devices. The state of every pair is computed and adjusted during the tuning phase, such that the average errors induced by the faulty devices are compensated. For the same case, this method reduces the accuracy drop to 0.3% for ConvNet and 3.2% for ResNet-18. Simulation results in Fig. 7a and 7b indicate that, for a sub-percent average accuracy drop, these two (low overhead) techniques enable tolerance of ~$1.5\times10^4$ ppm defective devices in ResNet-18 and ~$3\times10^4$ ppm faulty devices in ConvNet, both numbers >100× better than the initial resiliency.

# Discussion

The results presented in this paper establish strong predictions on the performance of analog neuromorphic networks in the presence of detrimental imperfections. Till now, research on this area has focused on commercially unscalable techniques such as in-situ or chip-in-the loop techniques. Other than that, most previous works study the impact of a single nonideality on redundant networks using small datasets solely based on the simulations or data from practically nonviable devices. This work performs a comprehensive characterization of major imperfections in the most prospective analog-grade memory devices. Characterization results are then harnessed to develop accurate device models, which are then incorporated to train and test two massive DNNs.

Our experimental work confirms that the synapse imperfections are major obstacles in the path of further progress of mixed-signal neuromorphic systems. We show that eFlash and $TiO_2$ memristors have excellent retention characteristics and tolerable static nonlinearity. Using the balancing technique methodology [42], which optimizes the tuning voltage for minimum error, we report only <0.4% and <0.1% accuracy drop for RM1 and RM2, respectively, after including the static nonlinearity model in the forward pass of the ResNet-18. Temperature variations intrinsically change the state of any analog synapse with similar trends to the case studies of this paper and dramatically impact the performance. A naively designed mixed-signal DNN could randomly behave when operating at 100 °C. We propose three modifications in the training (temperature-sweep batch training), circuit (*k*-reference batch normalization), and tuning (state optimization) for designing a reliable neuromorphic hardware that can operate in a wide temperature range. The incremental incorporation of these techniques enables a sub-percent accuracy drop even in a complex classification task such as ImageNet. The results in this work also confirm the trade-off between noise and energy consumption in memory arrays. Layer-wise SNR optimization algorithm, dynamic range optimization, and fine-tuning the conditioned networks are the three proposed techniques, all applied during the training phase, which lead to 2.5× to 9× improvements (for ResNet-18) in the energy spent on synaptic devices. The improved energy consumption is only due to energy saving in performing computations in the memory arrays and is not directly related to system-level performance. Note that such improvement is with respect to the baseline model with similar dedicated SNRs in each layer and no optimizations for improving noise performance. Note the general noise analysis (in Supplementary Fig. 4) shows this improvement could also be used to increase the throughput or dynamic range in various layers.

Further, this paper shows that the intrinsic defect tolerance of deep neural networks falls short in larger and more complex tasks: with >500 ppm defective devices, the accuracy drop increases drastically beyond 1%. For a mature technology like eFlash, the fault probability is well below this intrinsic range, while for the emerging passive ReRAM, the paper introduces two approaches, both applied during the tuning phase, to enhance the margin by a factor of >100×. The passive ReRAM technology offers the highest device density and monotonic 3D integration. However, the high uniformity requirement that enables analog tunability is the greatest challenge in fabricating these circuits. This paper shows that naïve tuning and training of a DNN model implemented with 64×64 kernels featuring 25% normalized switching threshold variations and (most promising results demonstrated to date in [37]) would suffer from a significant accuracy drop. We employ an advanced two-step programming algorithm during the tuning phase and a preprocessing step during the training phase that reduces this accuracy drop below 1% in ResNet-18. Note that since tuning is a very infrequent process in ex-situ trained memristive systems, the energy consumption during tuning is not a major concern.

Although the proposed approach is examined using two specific memory technologies, it is not tied down to particular features of these devices. Hence, this holistic approach could be applied in any mixed-signal neuromorphic implementations. For any memory technology, whether it is a FET-style synapse like eFlash or a resistive switching device similar to our memristive stack, imperfections may be modeled and included in the process of developing, training, and tuning the neuromorphic network. This study is also decoupled from the choice of a mixed-signal architecture in part because changing the structure of these massive networks in our simulation environment has a significantly destructive impact on the inference and training runtime of the model. Besides, the impact of the studied imperfections is expected to be the same in different architectures, and our holistic approach does not depend on a specific feature of the mixed-signal accelerators.

High-order nonidealities such as temperature dependency of static nonlinearity, noise, etc., are neglected in our simulations because they are far less impactful. Besides, although the proposed techniques are analyzed and simulated individually, they are entirely independent and could be applied together. Nevertheless, in many cases, imperfections devitalize each other, e.g., memristive devices become more linear and less noisy at elevated temperatures. The IR drop [45] is neglected in our study because it is nearly impossible to simulate its effect in large-scale neuromorphic systems. Ref. [42] proposes a bootstrapping method that effectively tackles it at the expense of monopolizing two CMOS metal layers. Ref. [26] uses an efficient conversion algorithm to mitigate the impact of IR drop, and Ref. [46] proposes to add an extra series resistance in peripheries to equalize the parasitic resistance seen by all the devices. The impact of endurance failure is not covered in this study since endurance requirements for ex-situ training of mixed-signal neuromorphic circuits are relaxed (e.g., $<10^5$) compared to in-situ approaches that rely on frequent write operations and most nonvolatile memories, including the demonstrated devices in this paper, can offer such specifications.

In our studies, we found that when no particular technique is used to mitigate imperfections, mapping 2 outperforms mapping 1 in terms of reliability at the cost of extra energy consumption. However, the proposed holistic approach allows us even to employ mapping 1 for weight to conductance conversion and saves extra energy that was previously inevitable. The most appealing feature of this approach is its scalability and the fact that it can be easily integrated into the design flow of these massive systems. The modifications performed in the training phase do not require any specific knowledge of imperfections (e.g., location of faulty devices) or individual chips and could be integrated with the typical ex-situ training procedure. The circuit modifications include additions of a simple temperature sensor circuit, low-cost hardware to support multiple batch normalization parameters per neuron, and an extra column in each kernel, with the total overhead that barely reaches 1% of an entire DNN chip. The state optimization and advanced tuning algorithms also do not require any extra hardware and are applied simply for every chip during the ex-situ tuning. Although our proposed holistic approach might slightly increase the training time, for the majority of the networks, the extra imposed training time is comparable with the training time of the baseline model, which is also negligible since training is performed only once in ex-situ trained systems and the developed model is used for a generation of deployed mixed-signal inference accelerators.

In conclusion, we have performed extensive characterization of imperfections in mainstream analog-grade synaptic devices and developed a holistic hardware-aware ex-situ approach to combat their detrimental impact on the performance of DNNs. Supplementary Fig. 13 compares this study with previous work and clearly validates the contribution of this work. The proposed approach includes modifications in training, circuit, state optimization, and tuning algorithm and has minimal areal or power overhead. Our methods are successfully tested on two large-scale deep neuromorphic networks. We believe that our results significantly improve the accuracy and efficiency of mixed-signal DNNs. Future research should focus on developing generalized device models to evaluate the effectiveness of our approach as a general solution and implementing the proposed methodology in fully-integrated neuromorphic circuits.

# Methods

## *Circuit Fabrication and Operation*

The 64×64 memristor crossbar consists of 4K passively integrated Al(70)/TiN(45)/$Al_2O_3$(1.5)/$TiO_{2-x}$(30)/Ti(15)/Al(90)/TiN(80) devices, with thickness shown in nm. Though the bi-layer ($Al_2O_3$/$TiO_{2-x}$) insulating material in such crossbars is similar to our previous works, the main difference is the etch-down patterning process that allows for attaining a higher aspect ratio and smoother electrodes, and consequently, improving *IV* uniformity and scaling up the circuit complexity. Besides, the low-temperature budget makes the developed fabrication process suitable for BEOL CMOS integration. First, Ti/Al/TiN metal stack is deposited on a 4-inch Si wafer with 200 nm of thermally grown $SiO_2$ using reactive sputtering. Then, ~250-nm wide bottom electrodes are patterned by deep ultraviolet lithography stepper with an anti-reflective coating and planarized by depositing 300 nm of $SiO_2$, smoothening it using a chemical-mechanical polishing process. Bottom electrodes are then opened by etch-back with CHF3 plasma before depositing the $Al_2O_3$/$TiO_{2-x}$ active switching bilayer through atomic layer deposition and reactive sputtering. Oxygen descum is no longer conducted after this deposition to keep $TiO_{2-x}$ stoichiometry. Then, top electrodes with Ti/ Al/TiN stack are deposited and patterned similar to bottom electrodes. Further, to suppress line-to-line leakages and open bottom electrode contacts, the switching layer outside the crossbar region is etched with CHF3 plasma. Finally, Ti/Au pads are formed to facilitate the wire bonding process, and thermal annealing is performed for 1 minute in $N_2$ gas with 2% $H_2$ at 350˚ C.

The eFlash chip includes an array of 12×10 redesigned split-gate memory array fabricated in Global Foundry's 55 nm LPe process. The modified array is 3× is larger than the original array. But it supports high-precision individual analog tuning of each cell, with <1% accuracy, while keeping the highly optimized cells, with their long-term retention and endurance, intact. Multiple chips are employed for characterization purposes throughout this paper.

Both memristor and eFlash chips are wire-bonded and mounted on a custom printed board for measurements, and electrical characterization is performed via Keysight B1500A parameter analyzer and B1530A measurement tool (see supplementary Fig. 11). Keysight 34980A and custom-made switch matrices steer the connections to memristive crossbar inputs/outputs and eFlash array, respectively. The parameter analyzer and the switching matrices are connected to a personal computer via a general-purpose interface and universal serial buses and controlled using a custom C++ code.

## *Forming, Tuning, and Operation of Memristive Devices*

Memristive devices require an electroforming process upon fabrication. An automated procedure to electroform devices is developed. We select a pristine device via the switch-matrix and apply an increasing amplitude current sweep while leaving other unselected devices and their corresponding electrodes floated. The low-voltage (0.1 V) conductance of the device is monitored after each pulse to check whether the device is formed or not and to avoid overheating it. If forming is failed for a device, the maximum current amplitude in a sweep is increased, and the process is repeated. Though the pulse width is fixed at 1 ms, the maximum current amplitude is increased automatically to overcome the additional leakage induced by the previously formed devices during the runtime. Whenever the algorithm fails to form a device (after achieving a certain threshold of pulse amplitude), the devices that share either top/bottom electrodes with the selected device and their conductances are more than 15 μS are reset. Then, the same forming process is repeated for the same memristor. Such leakage removal procedure significantly improves the $IV$ uniformity in the final formed crossbar. When forming fails for a device, it usually requires a higher forming current (supplied in future runs). In this situation, we switch to form the next device as the threshold is increased after each round of trying all devices in the crossbar. In a properly annealing crossbar, we can remarkably form >99% of devices in several rounds.

Upon forming the entire crossbar, we can tune and adjust the state of each device to a desirable conductance using the V/2 scheme and write-verify algorithm. Employing the switch matrix, we apply V/2 to a selected top (bottom) electrode and -V/2 to a selected bottom (top) electrode and ground others. After each pulse, the device is rested for 100 μs to discharge any volatile state (or charge) and before its low-voltage conductance is monitored. While the pulse duration is kept at 1 ms in our tuning algorithm, the pulse amplitude (V) is increased progressively from 0.5 V by 0.1 V steps (to speed up the tuning) until either the target tuning precision or the maximum pulse amplitude (2 V for set and -2.5 V for reset) is reached. The pulse polarity is alternated, and its amplitude is initialized when the device passes the target conductance. We consider two forms of mapping signed weights ($W$) to the conductance of a pair of memristive devices ($G_+$ and $G_-$). In the more power-efficient mapping 1, we use $G_\pm = G_{\text{min}} + \Delta G_{\text{max}}((|W| \pm W)/2|W|_{\text{max}})$ in which $G_{\text{min}}$ is the minimum conductance and $\Delta G_{\text{max}}$ is the conductance dynamic range, and $|W|_{\text{max}}$ is the maximum weight magnitude in a layer. In the more reliable mapping 2, we use $G_\pm = G_{\text{b}} \pm \Delta G_{\text{max}}(W/2|W|_{\text{max}}))$ in which $G_{\text{b}}$ is the midrange conductance. The temperature measurements are performed by heating the crossbar package using power resistors and setting the desired temperature by a feedback circuit controlled via Eurotherm PID temperature controllers. The system operates reliably with ±1 °C accuracy and high-speed response time. First, 350 random devices are tuned at room temperature to various random states. Then, we set the desired temperature in the PID circuit, wait for 5 minutes to ensure the die is sufficiently heated, and then record the state of each device in 0.1 V and 1 s intervals using WGFMU units. The entire process (including random tuning) is repeated 4 times. The accelerated retention measurements are also conducted by first tuning the devices at the room temperature and then using the same setup for baking the crossbar at 100°C for >25 hrs, and in-situ monitoring their states. We use linear regression to remove the setup noise from our measurements and estimate the conductance drift in room temperature using the Arrhenius law. We consider 500 memristors, each tuned to 6 random states in every measurement round. The state of each device is recorded at 0.1 V bias using 10-ms pulses in 400 s intervals after the chip is heated sufficiently. The static nonlinearity measurements are performed by selecting a device using the switch matrix and executing an $IV$ pulse sweep via B1500's SMU units with $V_{\text{max}}$=0.3 V, 5 mV steps, and 10 ms pulse width. Note that unselected electrodes are grounded, and hence, there is no sneak path current during the read command (similar to the inference phase). We record the $IV$ characteristics for 350 devices, each tuned to multiple states. The dynamic moded is developed by measuring the average switching characteristics of 500 devices in the crossbar using 2 ms pulse width with varying pulse amplitudes (0.5 V to 1.8 V for set; -0.5 V to -2 V for reset) and several initial conductance points (7.5 μS, 10 μS, and 25 μS for set; 16 μS, 25 μS, and 50 μS for reset). The tunning statistics are collected over the full crossbar after effectively programming all devices to desirable states (a grayscale quantized Einstein portray downsampled to 64×64 pixels) with 5% target tuning precision in the range of 10 µS to 100 µS and removing the defective devices. The relative tuning error is then obtained by measuring the device conductances at 0.25 V after tuning the entire crossbar in 3 rounds. The programming is performed similarly to previous experiments, and <1% defective devices are skipped to avoid disturbing other devices.

### *Tuning and Characterization of eFlash Devices*

The redesigned eFlash memory arrays consist of supercells that include 2 memory cells sharing a source line (SL). Other than that, each device has a wordline (WL), erase-gate (EG), control-gate (CG), and bit-line (BL). WLs, CGs, and SLs are routed in the same direction, while EGs and BLs are shared perpendicularly. Unlike passive memristor arrays, half-select inhibition in both programming and erasing are sufficient to eliminate disturbance, i.e., every device can be tuned individually without the need for disadjusting the rest. Like the passive crossbar, the write-verify algorithm is used to adjust a cell to a desirable state (the read current in a nominal biasing condition) with high precision. The state of an eFlash memory is increased (programmed) via hot-electron injection by applying a pulse to its source line while the selected row is biased at $V_{\mathrm{EG}}$ =5 V, and $V_{\mathrm{BL}}$ =0.8 V; unselected rows are biased at $V_{\mathrm{EG}}$ =0 V, and $V_{\mathrm{BL}}$ =2.5 V; the selected column is biased at $V_{\mathrm{WL}}=1.5$ V and $V_{\mathrm{CG}}=10$ V; and unselected columns are biased at $V_{\mathrm{WL}}=0$ V, $V_{\mathrm{CG}}=2$ V, $V_{\mathrm{SL}}$ =0.8 V. We can decrease (erased) the state of a device continuously as well via Fowler-Nordheim tunneling by applying a pulse to the EG of the selected device while keeping all ports grounded, except for the unselected columns which are biased at $V_{\mathrm{CG}}=8$ V. We also use the following nominal conditions to read the state of a device from its BL port using WGFMU units with 2 ms pulse width: $V_{\mathrm{WL}}$ =1.2 V, $V_{\mathrm{BL}}$ =1.2, $V_{\mathrm{CG}}$ =2.5, $V_{\mathrm{EG}}$ =0, and $V_{\mathrm{SL}}$ =0.

The more promising gate-couple topology [43] is considered for computing the relative weight change in the presence of imperfections and mapping the changes in state currents to weight values. In this topology, the state of each synaptic weight pair is determined by $I_{\mathrm{state}}/I_{\mathrm{max}}$ in which $I_{\mathrm{state}}$ is the state of the eFlash cell implementing the weight and $I_{\mathrm{max}}$ is the state of the peripheral cell, both in nominal biasing conditions. Similar to memristive devices, we study two forms of signed weights ($W$) to differential current mapping ($I_+$ and $I_-$). Here the same peripheral device is used to bias the differential pair, and we use $I_\pm = I_{\mathrm{min}} + \Delta I_{\mathrm{max}}((|W| \pm W)/2|W|_{\mathrm{max}})$ in mapping 1 and $I_\pm = I_{\mathrm{b}} \pm \Delta I_{\mathrm{max}}(W/2|W|_{\mathrm{max}}))$ in mapping 2. $I_{\mathrm{min}}$, $I_{\mathrm{b}}$ and $\Delta I_{\mathrm{max}}$ are minimum and bias currents, and the considered dynamic range, respectively. The temperature setup and measurement procedure are similar to those of memristors. We tune 100 devices to various states ($I$<100 nA) in deep weak inversion at room temperature and then use the PID circuit to record their states in the nominal biasing conditions and 1 s intervals. The entire process (including random tuning) is repeated 5 times for every device. The retention test is performed by tuning the devices to various states at room temperature and then monitoring and recording their states in the nominal biasing conditions every 100 s using 2-ms wide pulses while baking the package at 100°C for >6 hours. The experiments are terminated after 6 hours as no considerable drift is observed after this period. The nonlinearity characterization in eFlash devices is performed by measuring the static *IV* characterizations of 200 devices.

### *Neuromorphic Benchmarks*

Supplementary Fig. 12 shows the architecture of the neuromorphic benchmarks. The ConvNet model is based on Lenet-5 [44] architecture that includes 6 layers: Conv1, a convolutional layer with 5×5 filters and 65 feature maps; Pool1, a max-pooling layer of 2×2 regions; Conv2, a convolutional layer with 5×5 filters and 120 feature maps; Pool2, a max-pooling layer of 2×2 regions); FC1, a fully connected layer with 390 neurons; and finally FC2, a fully connected layer with 10 output neurons. Batch normalization [2] is applied after each non-pooling layer, and rectified linear is used as the activation function in all the layers. The CIFAR-10 dataset consists of 60k 32×32 color images in 10 classes, with 6k images per class. The model is trained with 50k images and tested on the remaining 10k images of the dataset. Standard data augmentation techniques such as zero-padding with two pixels, cropping a random 32×32 region, and performing random horizontal flipping of images are employed. No mean subtraction is performed (all input values are positive). We use ADAM optimizer [3], cross-entropy cost function, a batch size of 64, a learning rate of 0.001, and 220 epochs. Model initialization is performed following suggestions in [34].

The ResNet-18 implementation is based on the pre-trained model available at the official model zoo of the Pytorch. It includes 21+2 layers: a convolutional layer with 7×7 kernels and stride of 2, a max-pooling layer with 3×3 kernels and stride of 2, 4 convolutional blocks with residual connections, each including 4 convolutional layers based on 3×3 kernels and strides of 2 and 1, a 7×7 average-pooling layer with the stride of 7, and finally a 512×1000 fully-connected layer that provides the output prediction corresponding to 1000 classes. The network is tested on 50k images and trained on ~1.3M images for 150 epochs with a batch size of 256, the learning rate of 0.1 that is divided by 0.1 every 30 epochs (step scheduling), cross-entropy cost function, weight decay of 0.0001, and stochastic gradient descent optimization with a momentum of 0.9. The two models are trained using 32-bit floating-point precision on Nvidia Titan X GPUs, and the learned parameters achieving the highest test accuracy are used as the baseline model. During the mixed-signal simulation, we convert weights into device conductance/current, incorporate the developed models and techniques in the simulation platform and baseline architecture, and execute training and inference tasks. Note that we have not mapped the network into any mixed-signal architecture (e.g., see [38]) since simulating the targeted massive benchmarks within these (mixed-signal) architectures is incompatible and practically impossible with current GPU platforms and will make our results architecture-specific.

### *Temperature Variation Compensation Methods*

The first method is temperature-sweep batch training, in which we include the temperature model of synapses in the training process by considering a new hyperparameter called training temperature ($T_\theta$). Before running each forward pass of the training, we assume the model is ready for deployment in a chip that operates at an ambient temperature $T_\theta$ and converter all weights to their corresponding synaptic current values. Using the device model, we adjust the resultant synaptic current values in every step based on the training temperature value. The altered synaptic currents are converted back to the equivalent software weights before the forward pass is executed. Triangular scheduling of the training temperature is adapted, i.e., $T_\theta$ is swept from 25 °C to 95 °C and vice versa by 10 °C steps in every batch.

The second method is executed by considering *k* reference temperatures with temperature-unique batch normalization parameters. Owing to the monotonic change in the statistics of preactivations (i.e., the shift and stretch of the preactivations) with respect to the temperature (Fig. 4a-b), a temperature-dependent correction signal allows us to minimize the induced error. Since generating such neuron-specific signals with adjustable temperature-dependency are costly, we use a quantized version of it through multiple batch normalization weights that effectively shift and scale preactivations. After the model is trained with the first approach, we find *k* reference batch normalization parameters by retraining it in a single epoch with a learning rate of 0.001 in *k* reference temperatures. During the inference, the temperature of the chip is sensed by a low-cost on-chip sensor and used to determine the proper batch normalization parameters that correct the distributions.

The state optimization approach is the third technique that mitigates the accuracy drop in a wide temperature range. Here, the mapping parameters are optimized individually for every weight targeting the lowest weight error across the full temperature range. Such design parameters are often selected to minimize the power consumption in eFlash memories or maximize the dynamic range in memristors. However, these design parameters are not necessarily the most optimum in terms of temperature variations and reliability. Since this approach comes with power consumption addition or dynamic range reduction, a methodology that finds quasi-optimal design points in either weight-conductance mapping functions is developed. Supplementary Fig. 7 numerically analyzes experimental data and power-accuracy trade-off and shows how we obtained the quasi-optimum design parameters for each device stack and weight mapping functionality.

### *Optimization Techniques for Mitigating Noise*

In the layer-wise SNR optimization algorithm, a fixed energy budget, equally distributed among all layers, is considered, i.e., all layers are initially assigned a fixed energy scaler. A simple heuristic approach is developed to increase the energy scaler in some layers and decrease it in others, such that the total computational energy is kept constant, but the accuracy gets improved in every step of the process. In our analysis, we assume $E_l \propto m_l/\rho_l$, in which $m_l$ is the number of operations in the $l^{\text{th}}$ layer, and $\rho_l$ is its assigned energy scaler. Also, $\Delta d_l^{\pm}$and $\Delta E_l^{\pm}$ are the changes in the inference accuracy drop (the noise sensitivity of $l^{\text{th}}$ layer) and energy rate of a layer, respectively, as a result of multiplying its energy scaler by $a^{\mp 1}$. Constant $a > 1$ is optimized empirically for every network. The idea is to ascertain light layers that offer the highest accuracy gain per energy and increase their SNR and then reduce the SNR in power-hungry and insensitive-to-noise layers such that the net change of energy consumption is fixed.

The following three-step procedure is repeated until no further improvement is obtained. First, we exclusively find the noise-sensitivity and energy rates of each layer, i.e., $\Delta d_l^{\pm}$ and $\Delta E_l^{\pm}$. Two simple cost functions are adopted to select a layer that provides the energy boost and share it among the other layers. 1) Targetting the least noise-sensitive layers that create the highest energy boost, we scale the energy scaler of a layer that maximizes $c_l^- = \Delta E_l^- / \Delta d_l^-$ by $a^{-1}$ (for other layers, $\rho$ remains unchanged at this step). Note that the higher the $c_l^-$, the more energy boost it gives per minimum loss of accuracy. Hence, this step effectively targets a layer that can share the maximum energy with others while leading to the minimum accuracy reduction. 2) We compute $\Delta E_{l,\text{x}}^+ = (\Delta d_l^+ / \sum \Delta d_l^+)\text{Log}_{10}(\sum \Delta E_l^+ / \Delta E_l^+)$ for every layer and normalize it to find the fair share of every layer from the provided boost $\Delta E_l^-$ , i.e., $E_l^{\text{n}} = E_l^{\text{o}} + \Delta E_l^-(\Delta E_{l,\text{x}}^+ / \sum \Delta E_{l,\text{x}}^+)$ where $E_l^{\text{n}}$ and $E_l^{\text{o}}$ are the updated and previous energy budget of $l^{\text{th}}$ layer, respectively. The intuition is to provide a higher energy share to the layers that boost accuracy more while requiring less energy. 3) The energy scaler of each layer is updated using $E_l \propto m_l/\rho_l$ and the results are validated by simulating the noisy model using the new assortment of energy scalers. If the accuracy is decreased, step 1 is repeated using the second layer in $c_l^-$ ranking, and so on.

The second technique is a progressive brute force search of semi-optimized signal ranges. The optimized accuracy and the signal ranges from the previous method are used as references. We consider the following list of clipping threshold percentages: 2, 4, 5, 7, 10, 20, 30, and 40. Starting from the smallest threshold, we diminish the range of activation signals of a layer with the threshold and verify the impact by running the inference test. If the validation accuracy is better than the initial reference accuracy, the signal range is updated. The algorithm is executed in 3 rounds, after which reducing the signal ranges often leads to no further improvements. The third approach is to fine-tune the conditioned network parameters such that it can adapt to the signal and noise statistics of the physical hardware. For ResNet-18, the pre-trained network is trained for extra 25 epochs with noise included in the forward pass of the simulation, using a manually-optimized learning rate of $10^{-4}$ and weight decay of $10^{-4}$.

### *Defect-Tolerant Techniques*

The first approach to improve the defect tolerance is to retune the pair memory device of a defective device and minimize the weight mapping error. When $G_{\pm}$ is stuck to $G_{\max}$, we use $G_{\mp} = G_{\max} \mp \Delta G_{\max}((W \pm |W|)/2|W|_{\max})$ for pair-device retuning. When $G_{\pm}$ is stuck to $G_{\min}$, $G_{\mp} =$ $G_{\min} + \Delta G_{\max}((|W| \mp W)/2|W|_{\max})$ is used to retune the paired device. When $G_{\pm}$ is stuck to $G_{\min} < G_{\mathrm{x}} < G_{\max}$, $G_{\mp} = G_{\mathrm{x}} \mp \Delta G_{\max}(W/|W|_{\max})$ is (clipped to $G_{\max}$ or $G_{\min}$, if needed) used to retune the paired device. Clearly, if two defective devices constitute a synapse, it is not feasible to compensate for its weight mapping error. Besides, the limited dynamic range of pre-activation makes them susceptible to a small constant shift in a synapse output. The second method alleviates this issue by compensating for such shifts through an extra pair of analog memories (single column) per neuron per processing kernel. Such devices are always driven by a fixed maximum range signal, and their states are adjusted during the tuning phase to minimize the average shifts. To automate the procedure, after the pair-wise tuning is performed on a defective model, we use a tiny part of the training set ~7k and 1.5k images in ResNet-18 and ConvNet, respectively, to recompute the biases in batch normalization layers (that simply shift the pre-activation signals) and find the conductance of extra devices.

## Acknowledgments

This work was supported in part by a Semiconductor Research Corporation (SRC) funded JUMP CRISP center and in part by NSF/SRC E2CDA grant 1740352.

## Contributions

M. R. M., Z. F., and D. B. S. conceived the idea. Z. F. performed the simulations. M. K prepared the codes for the baseline networks. M. R. M. performed the experiments and developed the models. H. N. and H. K. fabricated the memristor crossbar. M. R. M. and H. N. built the experimental setup. M. R. M. wrote the manuscript. All authors discussed the results.

## Competing Interests

The authors declare no competing financial interests.

## Data Availability

The data that support the plots within this paper and are available from the corresponding author upon reasonable request.

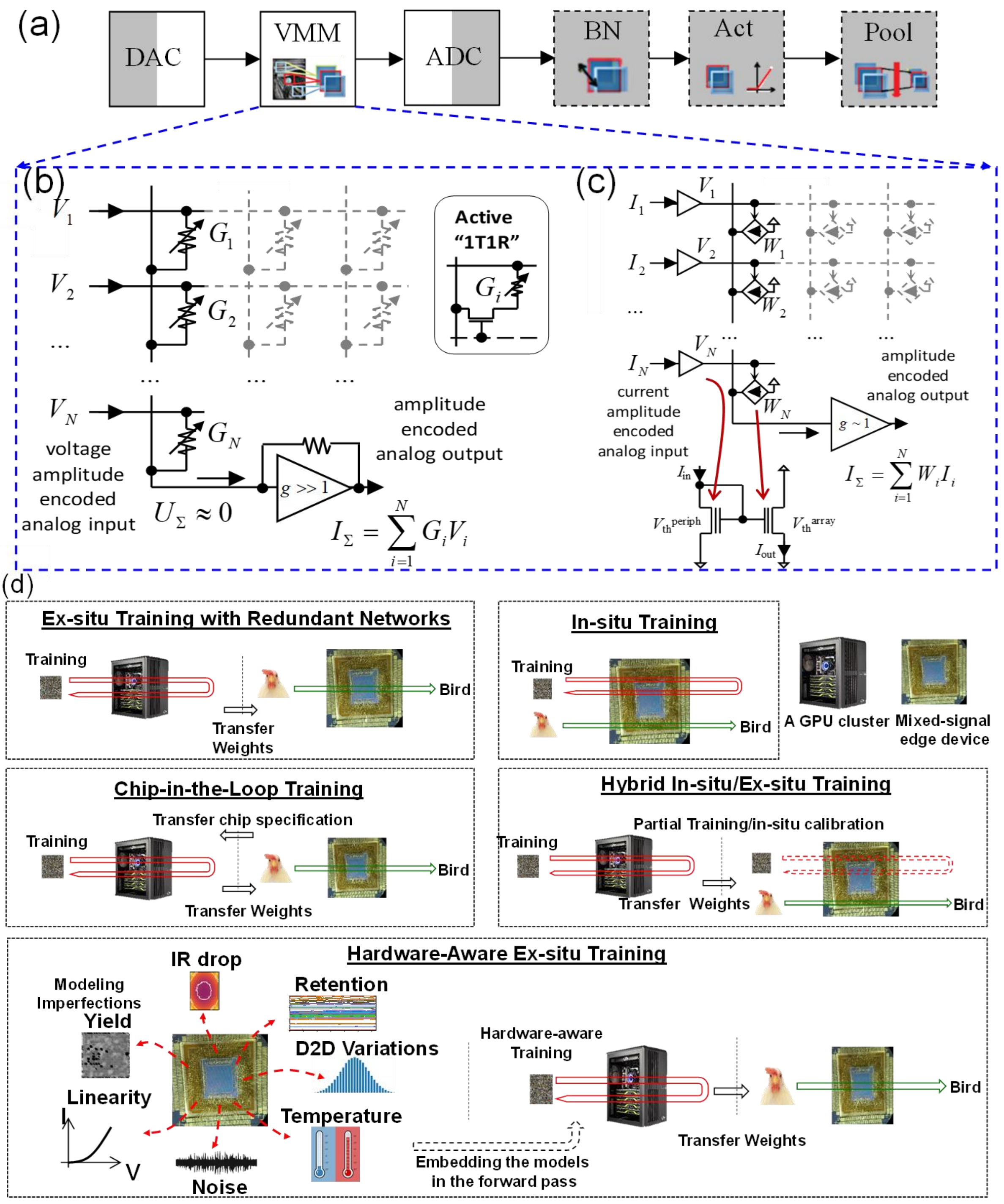

**Figure 1. Mixed-signal neuromorphic circuits.** (a) Major computations involved in every layer of a mixed-signal implementation of a modern neuromorphic classifier. Highlighted in gray (white) are those typically implemented in the digital (analog) domain (DAC: digital to analog converter, VMM: vector-by-matrix multiplier, ADC: analog-to-digital converter, BN: batch normalization, Act: activation function, Pool: an optional pooling layer, FC: fully-connected classifier). VMM implementation using (b) memristive crossbars and (c) gate-couple eFlash memory arrays. Network weights are encoded into two the conductance of memristors (synapses) or the ratio of the state currents of two eFlash devices to a peripheral eFlash memory (see the method section for the mapping functions). The input/outputs are often encoded as voltages ($V_i$) in memristive circuits and currents ($I_i$) in eFlash VMMs. (d) Various

approaches to mitigate imperfections in mixed-signal neuromorphic networks. Specifically, we focus on hardware-aware ex-situ training in which the imperfections are initially characterized and modeled. Training algorithms are then modified with imperfection models included in the forward pass of training mode to make the ex-situ trained model resilient toward them in the test phase.

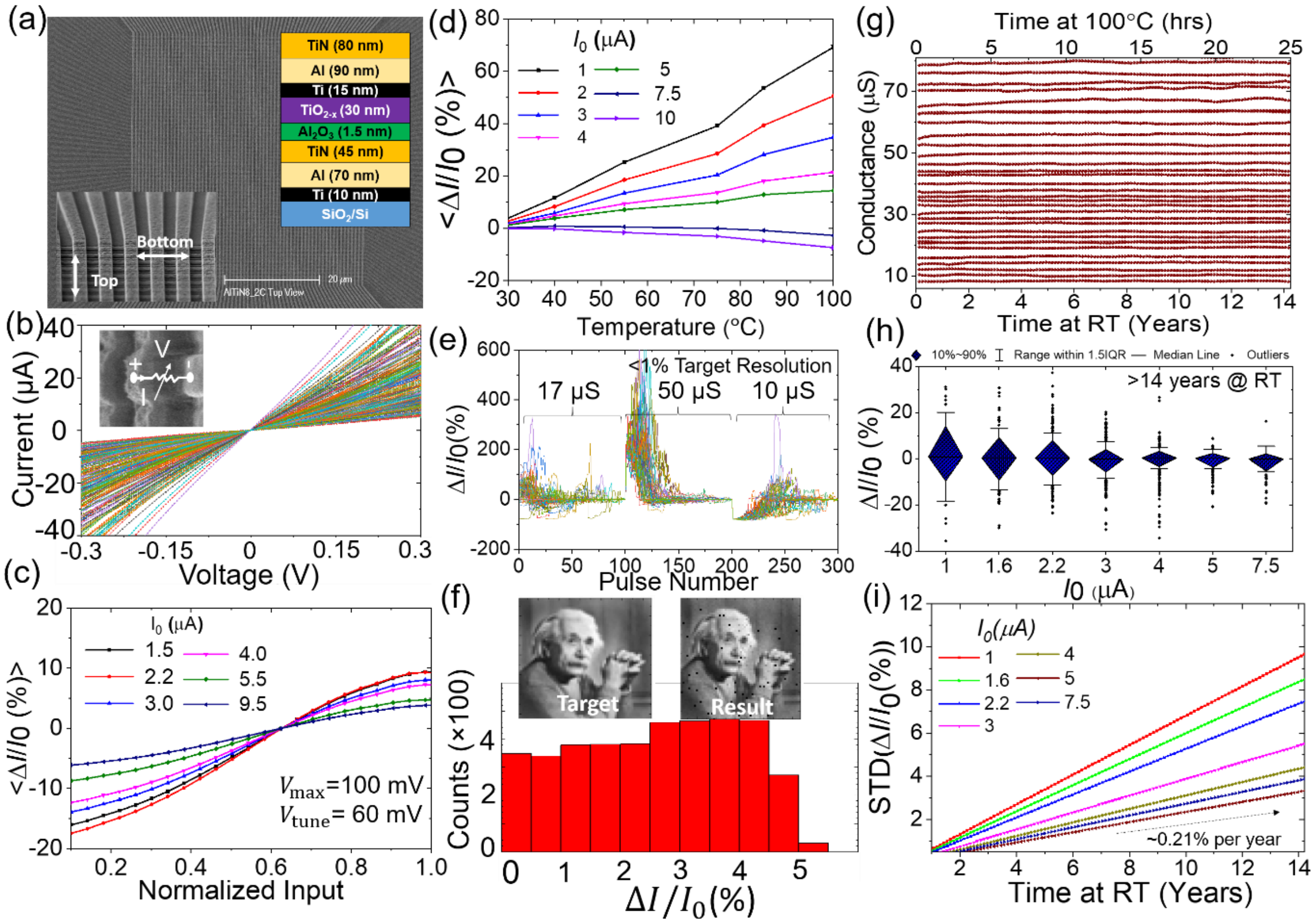

**Figure 2. Memristor crossbar characterization.** (a) SEM image of the full 64×64 memristor crossbar array. The bottom left, and bottom right insets show material layers at the device cross-section with corresponding thicknesses in nanometers and zoomed-in to a portion of the crossbar, respectively. (b) Low-voltage *IV* characteristics of 350 devices programmed to various states. The inset figure shows the cross-section of a device and how a voltage is applied across the device during a low-voltage read operation. (c) The average relative static nonlinearity error in memristor synapses $\Delta I/I_0 \times 100$ for the same 350 devices, tuned in various states. Static nonlinearity is computed with respect to the measured state current at the tuning voltage ($V_{\text{tune}}$). The static nonlinearity error is obtained assuming 60 mV reference tuning voltage and 0.1 V maximum voltage. (d) The average relative change in state current versus temperature for 350 memristive devices tuned to various states. Panel (e) shows how relative error changes when devices are tuned using a write-verify algorithm with <1% target relative error. The results are provided only for 50 random devices for clarity. (f) The final tuning error distribution in 64×64 crossbar after all devices are tuned. The desired device conductances in the range of 10 µS to 100 µS, which corresponds to the grayscale quantized Einstein image and their actual measured values after completing tuning with 5% target error. <1% non-switchable devices are excluded for clarity [37]. (g) Stable analog operation after >25 hours of baking the memristor crossbar in 100 °C that translates to >14 years of room temperature operation, assuming conservatively 1.1 eV activation energy. (h) The distribution of relative retention loss error ($\Delta I/I_0 \times 100$), where $I_0$ is the initial sensed current for 400 memristors, each tuned to 7 random states after projected 14 years of room temperature operation. Panel (i) shows the corresponding standard deviation of the relative conductance change versus time binned to different conductances. The conductance is measured at 0.1 V in these experiments.

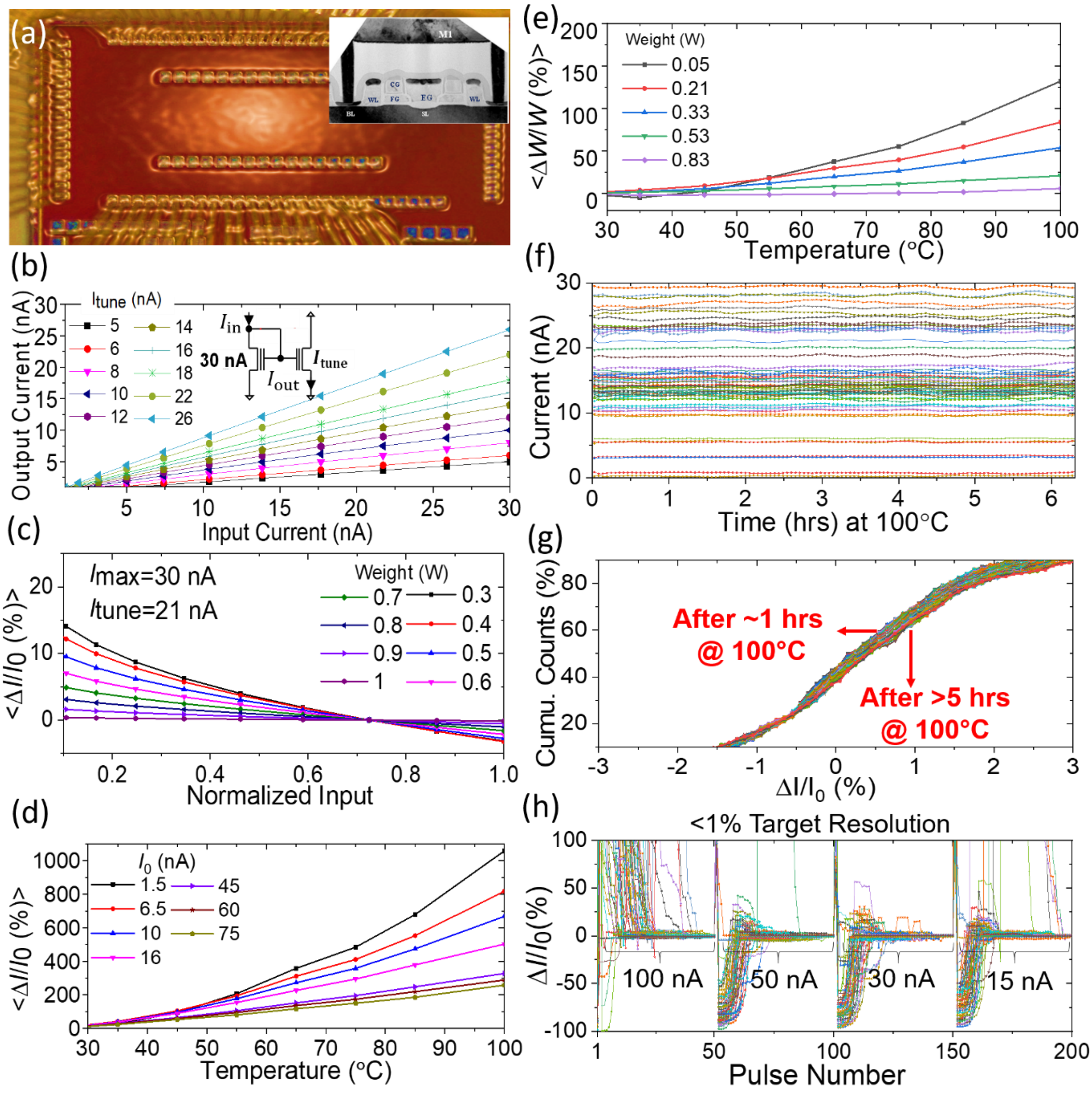

**Figure 3. Redesigned analog-grade eFlash characterization.** (a) Micrograph of the fabricated 12×10 eFlash array in Global Foundries' 55 nm CMOS process. The inset shows the TEM cross-sectional image of a supercell that includes 2 eFlash devices. (b) The average static input/output characteristics of 200 gate-coupled synapses given a peripheral cell tuned at 30 nA for various weight values and (c) the corresponding average nonlinearity error. In panel (c), the devices are tuned to the desired state (or weight) at $I_{tune} = I_{in}$=21 nA. Panels (d,e) show the average relative change in current measured using 100 devices and, correspondingly, the average relative change in the synaptic weight (assuming $I_{max}$=30 nA) of the gate-coupled structure versus temperature, respectively. (f) Accelerated retention test for 100 eFlash devices tuned in 5 different states measured at 100°C and nominal tuning conditions. Panel (g) shows the trend in the cumulative distribution function of the relative change in the current (@100 °C) for these devices. The relative change is within 1% for the majority of the devices. Panel (h) shows high precision tunability (<1% target relative error) in 50 analog-grade redesigned eFlash memories, tuned to various target states, each for a maximum of 50 pulses.

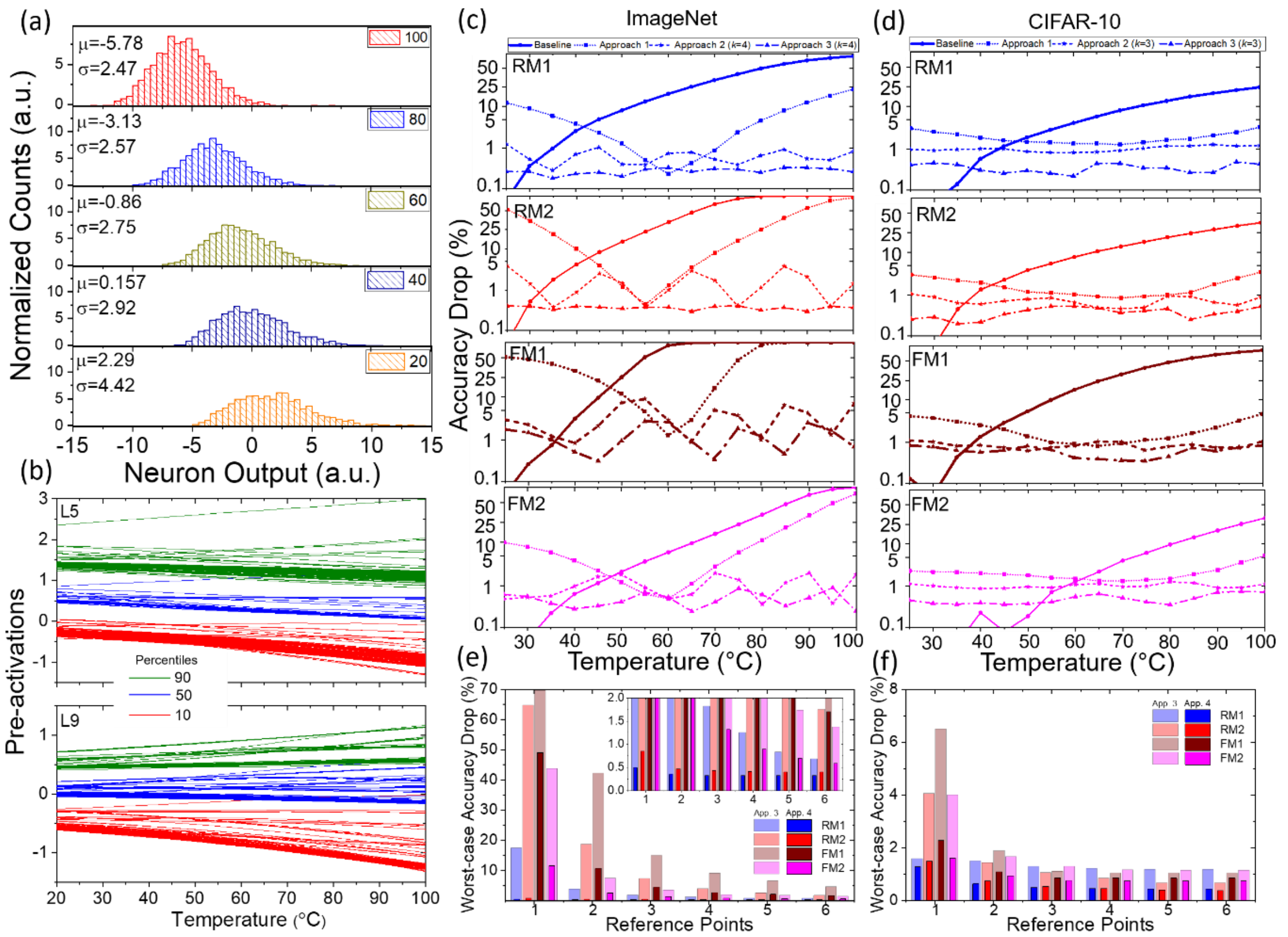

**Figure 4. Temperature compensation in mixed-signal DNNs.** (a) The distribution of the first neuron's output activations in the fully-connected layer of ResNet-18 for 10 inference batches in several temperatures (FM2). (b) The temperature dependency of 10, 50, and 90 percentiles of the pre-activation distributions (100 batches) of 100 random neurons in 2 different layers (RM1). (c,d) The accuracy drop versus temperature in ImageNet benchmark when using various synapse options to implement ResNet-18 model. Baseline corresponds to the evaluation case without any mitigation technique, while approaches 1, 2, and 3 refer to temperature-sweep batch training, *k*-reference batch normalization, and state optimization methods. These techniques are applied incrementally on the network. In approaches 2 and 3, we use 4 reference points. Panel (e,f) shows the worst-case accuracy drop in the 20-100 °C temperature range versus the number of reference points for approaches 2 and 3. Note that unlike panels (c,d), we use a linear-scale of the y-axis in panels (e,f) to emphasize the achieved improvement further. The inset shows the zoomed-in view of the low drop portion of the figure. Panel (f) shows similar results for the ConvNet model and CIFAR-10 dataset. RM1: ReRAM, mapping 1; RM2: ReRAM, mapping 2; FM1: eFlash, mapping 1; FM2: eFlash, mapping 2.

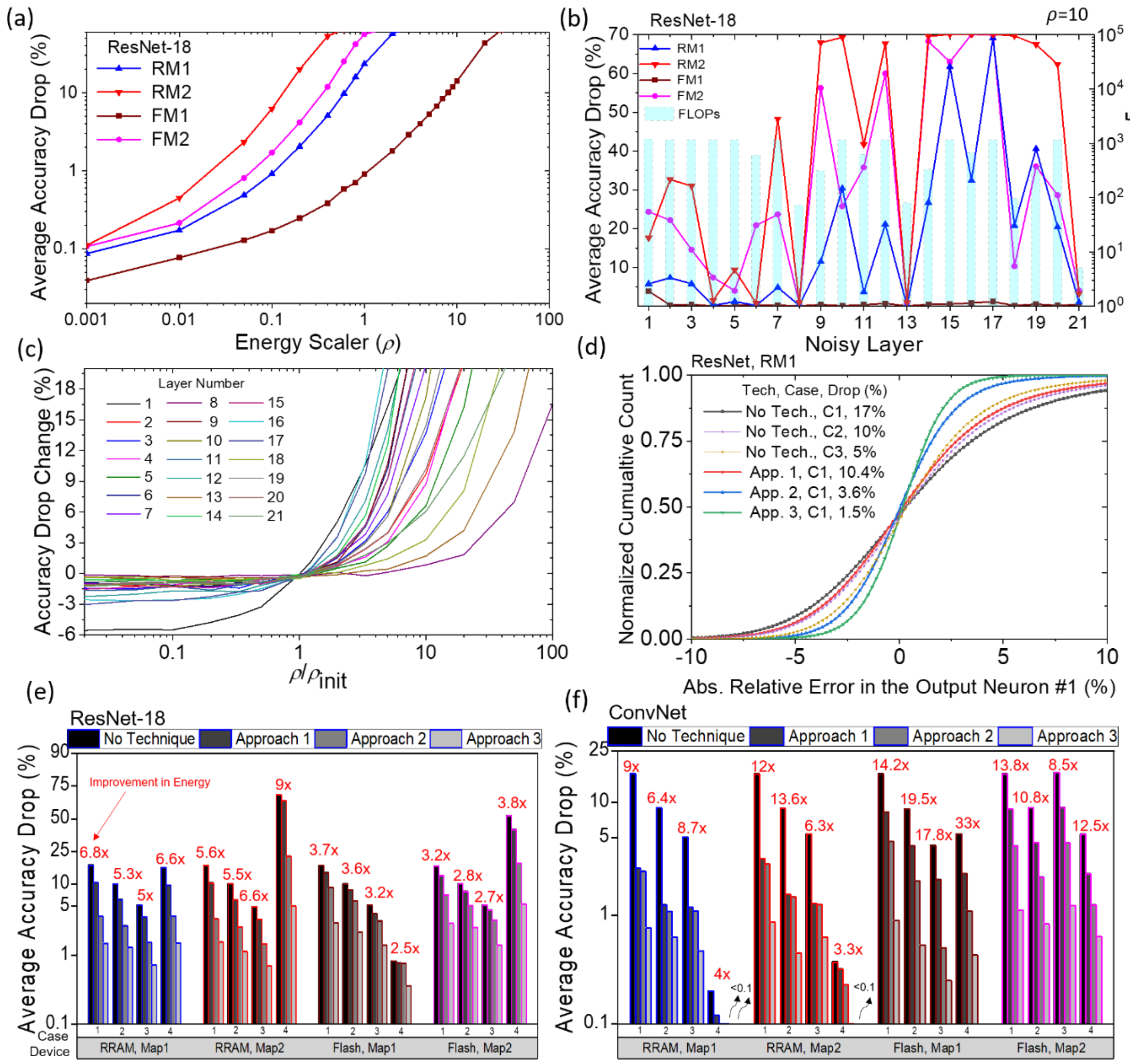

**Figure 5. Analog noise simulation results.** (a) The average accuracy drop in ResNet-18 versus the global energy scaler (the same energy scaler is used in all layers) for RM1, RM2, FM1, and FM2. (b) FLOPs (floating-point operations) in every layer and the average accuracy drop in ResNet-18 for various devices when only a selected layer is noisy. The energy scaler in the selected layer is set to 10. (c) The sensitivity of the accuracy drop in ResNet-18 (RM1) to the change in each layer's energy scaler. For each curve, only the energy scaler of a selected layer is altered. The energy scaler of other layers is fixed at $\rho_{\text{init}}$=1. (d) The cumulative distribution of the absolute relative error in the output neuron #1 of the ResNet-18 model (RM1) for several cases. Panels (e) and (f) show improved accuracy and energy efficiency for multiple cases in ResNet-18 and ConvNet, respectively. Note that here we only consider the energy spent in synapses, and the techniques are applied incrementally, i.e., Approach 3 includes our three proposed methods.

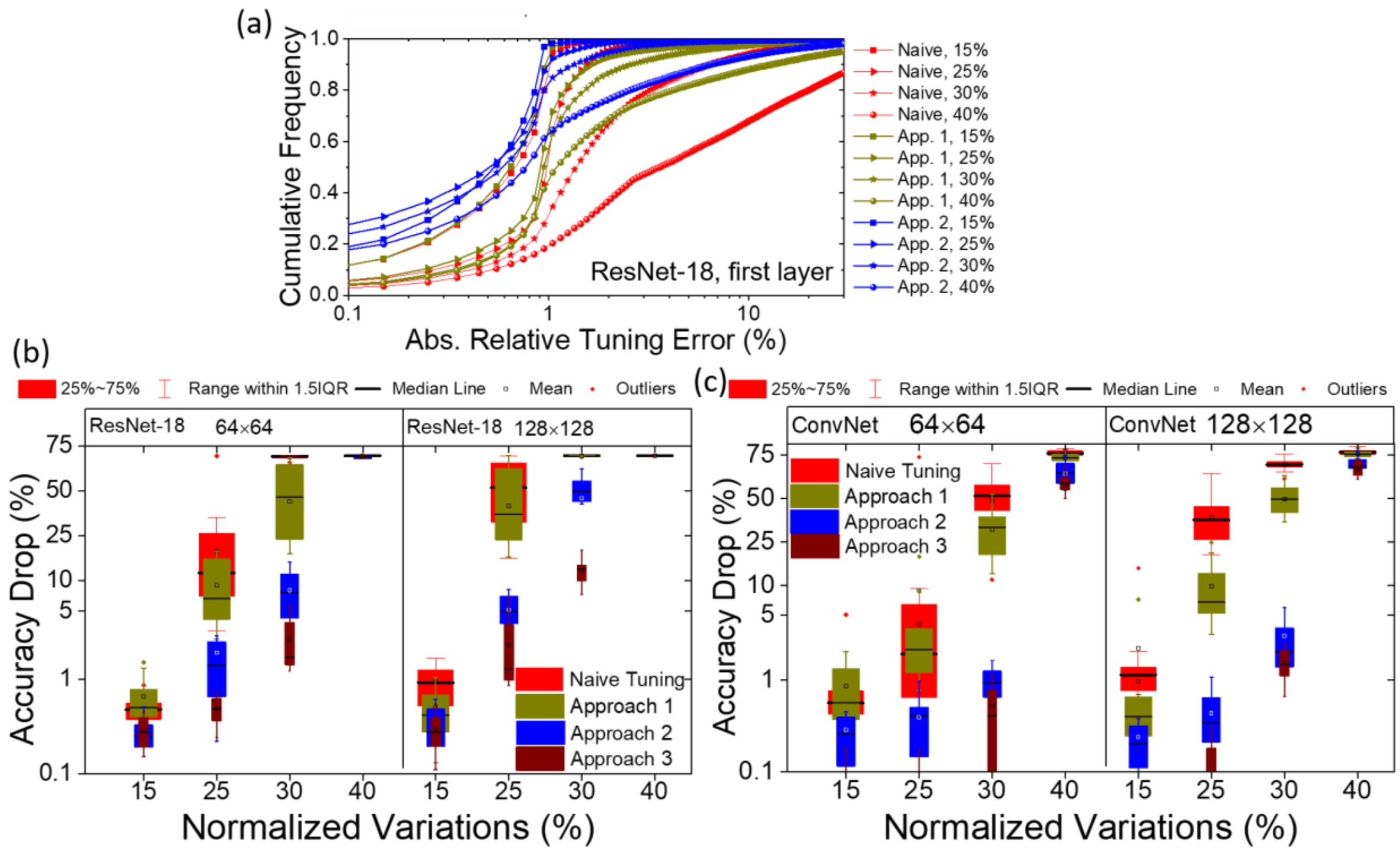

**Figure 6. Tuning precision simulation results.** (a) The cumulative distribution of the absolute relative tuning error ($\Delta G/G_0 \times 100$) for the programmed weights in the first layer of ResNet-18 (~2×9.5k devices) for various methods and normalized switching threshold variations after the 10th round (64×64 crossbars). The accuracy drop in 12 different tuned model instances in (a) ResNet-18 and (b) ConvNet based on various normalized switching threshold variations (15%, 25%, 30%, 40%) and crossbar sizes (64×64 and 128×128). For each data point, the ex-situ training process of the entire model is simulated before validating the model in the inference. The dynamic model and the naïve tuning method are discussed in detail in Supplementary Fig. 5. In approach 1, each crossbar is initially tuned using the naïve method for one round, and then the following thresholds are used to restrict the maximum write voltage in the remaining 9 tuning rounds: 2.2, 0, 2.1, 1.7, 1.5, 1.3, 1.1, 0.9, and 0.7 for the set operation, and 0, 2.2, 2.1, 1.7, 1.5, 1.3, 1.1, 0.9, and 0.7 for the reset operation. In approach 2, besides employing the previous technique, we initially set (reset) devices whose set (reset) switching thresholds are > 1.5 V and then exploit the feasibility of encoding a weight with various conductances in the differential pair implementation.

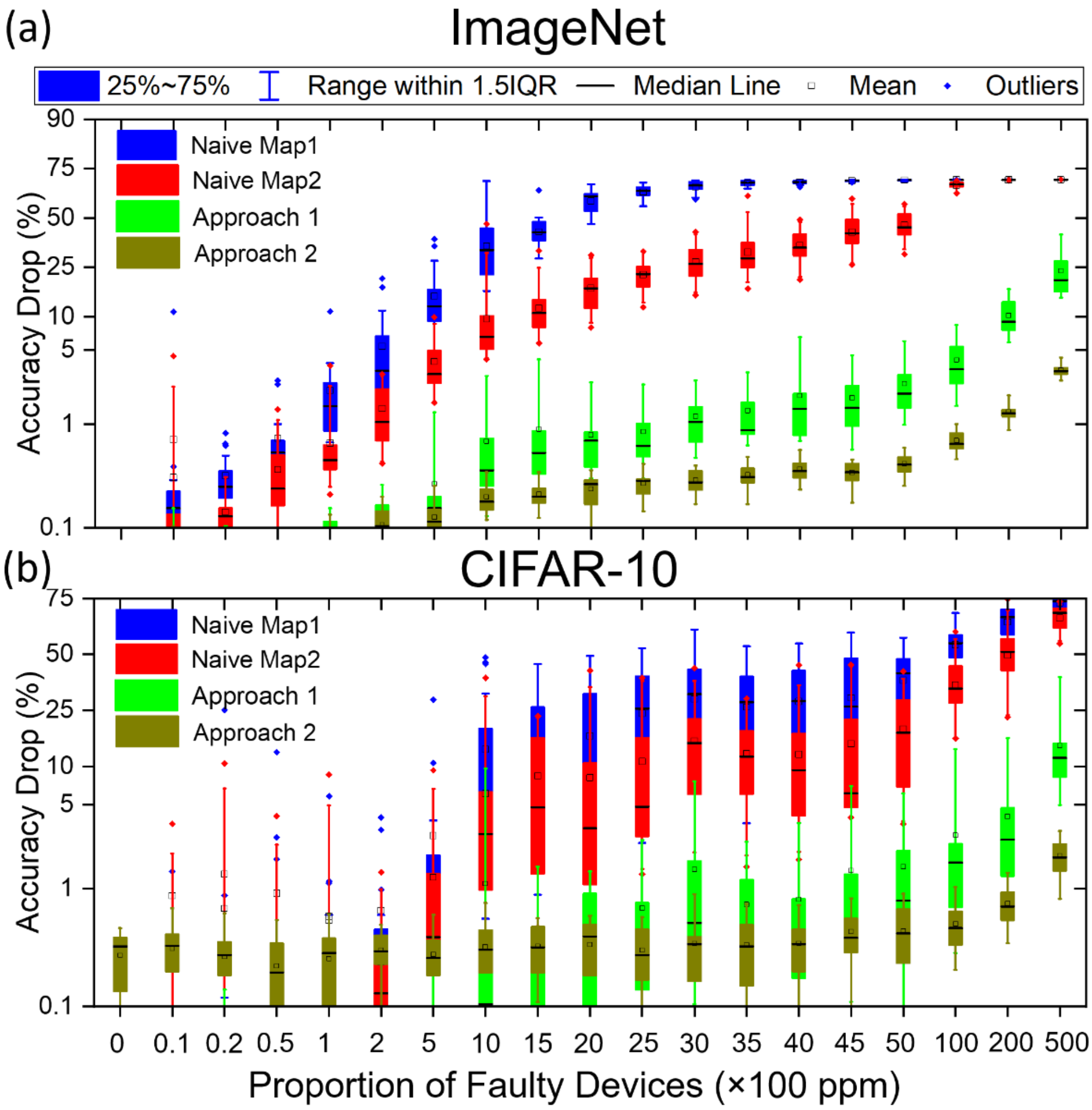

**Figure 7. Defect-tolerance simulation results.** Defect-tolerance improvements in (a) ResNet-18 and (b) ConvNet using the two incrementally applied approaches. For every point, the statistics are obtained over 20 runs. For each point, the same percentage of stuck at high conductance, low conductance, and random states devices are considered (see Supplementary Fig. 10 for the separate case studies). The locations of defective devices in each run are chosen randomly. The results of approach 1 (and approach 2) are the same for both mappings because of using the same compensation scheme in pair-wise adjustment (i.e., independent of the original mapping scheme).

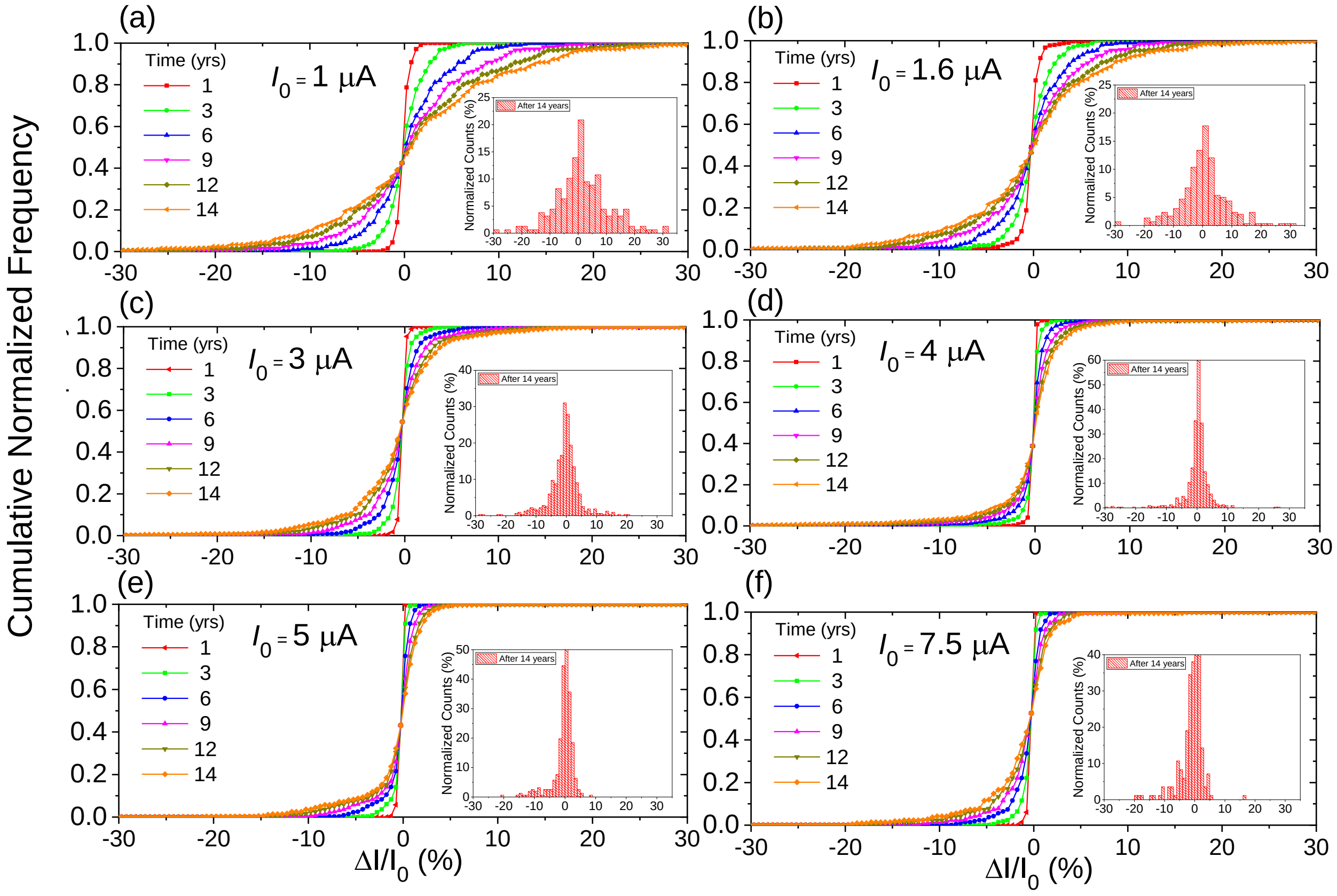

**Supplementary Figure 1. Extended measurement results of accelerated retention test in memristive devices.** Panels (a-f) show the cumulative normalized frequency of relative retention loss error among 400 devices tuned to various states. Accelerated retention tests are performed at 100°C at 0.1 V for more than >25 hours. The results are then projected to room temperature using the Arrhenius equation and 1.1 eV activation energy. The insets show the histogram of the error for the case of 14 years. Our results indicate that the retention loss is a bidirectional process for most devices and analog intermediate states, particularly midrange conductances. Note that moving towards high conductive states (e.g., panel (f)), we observe a trend that corroborates devices' tendency to move toward midrange conductances. In fact, we expect and observe a unilateral retention loss behavior upon hard switching devices to the extreme regimes (<5 μS and >150 μS). Nevertheless, the bilateral trend of retention loss of analog states is a positive feature since the tiny retention-induced errors average becomes even smaller when they average out in large matrix multiplier kernels.

(a) $(\Delta w/w_0) \times 100 = (T - T_0)(p_{00} + p_{10}w_0{}^{-1} + p_{20}w_0{}^2 + p_{30}w_0{}^3)$

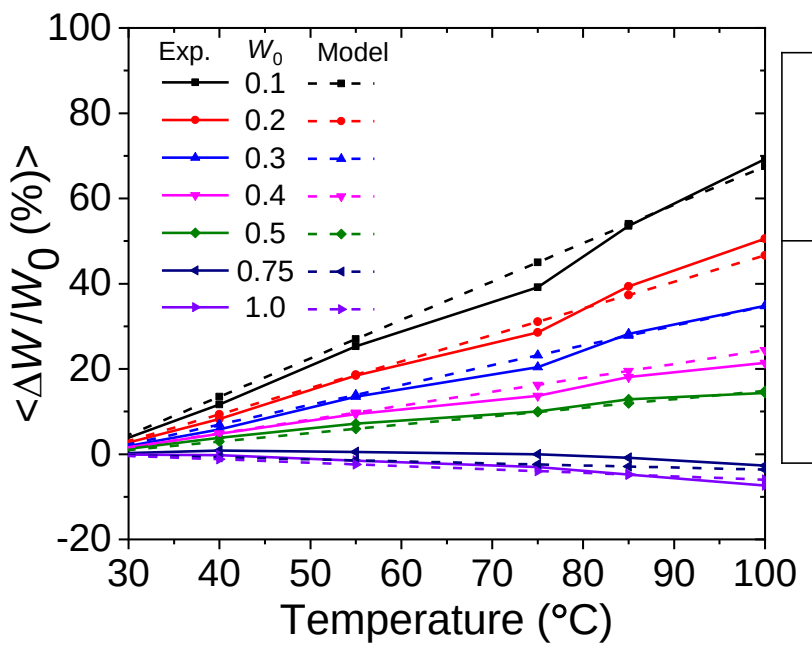

| | | |
|---|---|---|
| Params | $p_{00}$ | 0.48 |
| | $p_{10}$ | 0.44 |
| | $p_{20}$ | -0.02371 |
| | $p_{30}$ | 0.001765 |
| Goodness of fit | SSE | 70.57 |
| | R – Square | 0.9975 |
| | Adjusted R – Square | 0.9973 |
| | RMSE | 1.266 |

(b) $(\Delta w/w_0) \times 100 = (T - T_0)(p_{00} + p_{10}w_0 + p_{01}T + p_{20}w_0{}^2 + p_{11}T \times w_0 + p_{21}w_0{}^2T)$

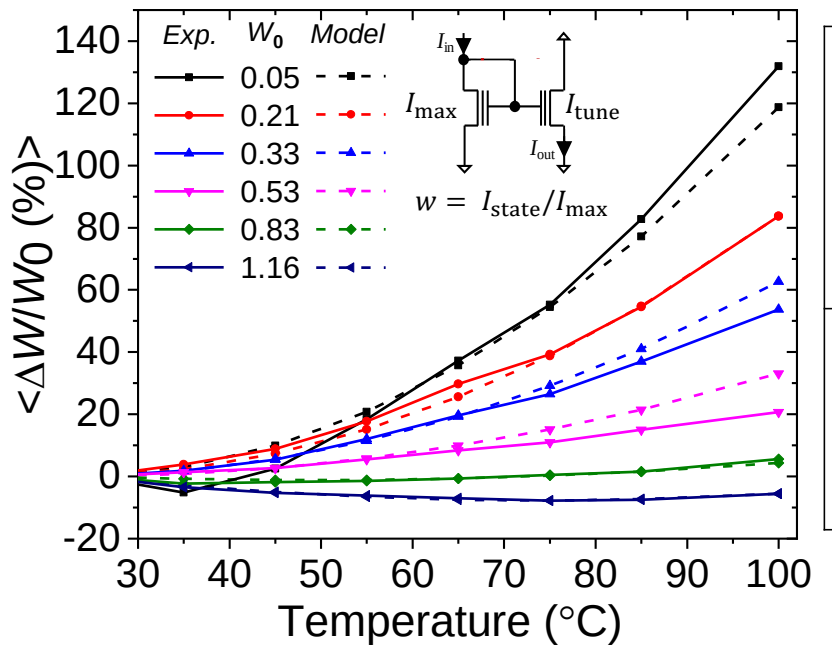

| | | |
|---|---|---|
| Params | $p_{00}$ | -0.4465 |
| | $p_{01}$ | 0.02185 |
| | $p_{10}$ | 1.121 |
| | $p_{11}$ | -0.04293 |
| | $p_{20}$ | -0.9239 |
| | $p_{21}$ | 0.02311 |
| Goodness of fit | SSE | 16.05 |
| | R – Square | 0.9811 |
| | Adjusted R – Square | 0.9788 |
| | RMSE | 0.6181 |

**Supplementary Figure 2. Temperature modeling in memristors and eFlash synapses.** The modeling results for analog-grade (a) memristors and (b) eFlash memories. Instead of using complex physics-based models that would significantly slow down the simulation time in our massive neuromorphic benchmarks, we use multi-order polynomial functions that decently and efficiently predict devices' average behavior. In both cases, the most optimum polynomial function is found manually by an exhaustive brute force search, and nonlinear least squares optimization with a trust-region algorithm is applied to find the optimum fitting parameters. To study temperature variations, the relative change in the weight $(\Delta w/w_0) \times 100$ of every device in a synaptic pair is modeled using $(T - T_0)(p_{00} + p_{10}w_0{}^{-1} + p_{20}w_0{}^2 + p_{30}w_0{}^3)$ for metal-oxide memristors and $(T - T_0)(p_{00} + p_{10}w_0 + p_{01}T + p_{20}w_0{}^2 + p_{11}T \times w_0 + p_{21}w_0{}^2T)$ for eFlash memories in which $w_0$ is the measured weight at nominal biasing conditions and $T_0 = 25$, $T$ is the die's temperature in Celsius, and $p_{ij}$s are the fitted parameters. The fitting results show excellent goodness of fit across the temperature range for both synaptic device candidates. In panel (a), a weight exactly corresponds to a device conductance (in a synaptic pair and μS), i.e., $W_0 = 0.1$ and $W_0 = 1$ correspond to $G_{\min}$ and $G_{\max}$, respectively. In panel (b), a weight corresponds to a device state (in a synaptic pair) over the peripheral device state, i.e., $W_0 = 0$ and $W_0 = 1$ correspond to $I_{\text{state}} = 0$ and $I_{\text{state}} = I_{\max}$, respectively. Since the peripheral state is often tuned at $I_{\max}$ (that is 30 nA in this figure), $W_0$ equalizes to the normalized weight. Note that in the case of eFlash memories, the model parameters change when a different $I_{\max}$ is used. Note that most synaptic devices exhibit similar trends, and we expect that similar modeling formats would be applicable to other devices as well. Only model parameters would be different. High-order nonidealities such as temperature dependency of nonlinearity, noise, etc., are neglected in our simulations because they are far less impactful, and they typically devitalize each other, e.g., they become more linear and less noisy at elevated temperatures. Hence, we neglect them in our modeling here.

$$\Delta y = x(x - x_{\mathrm{tune}})(p_{01}x + p_{03}x^3 + p_{10}w + p_{20}w^2 + p_{30}w^3)$$

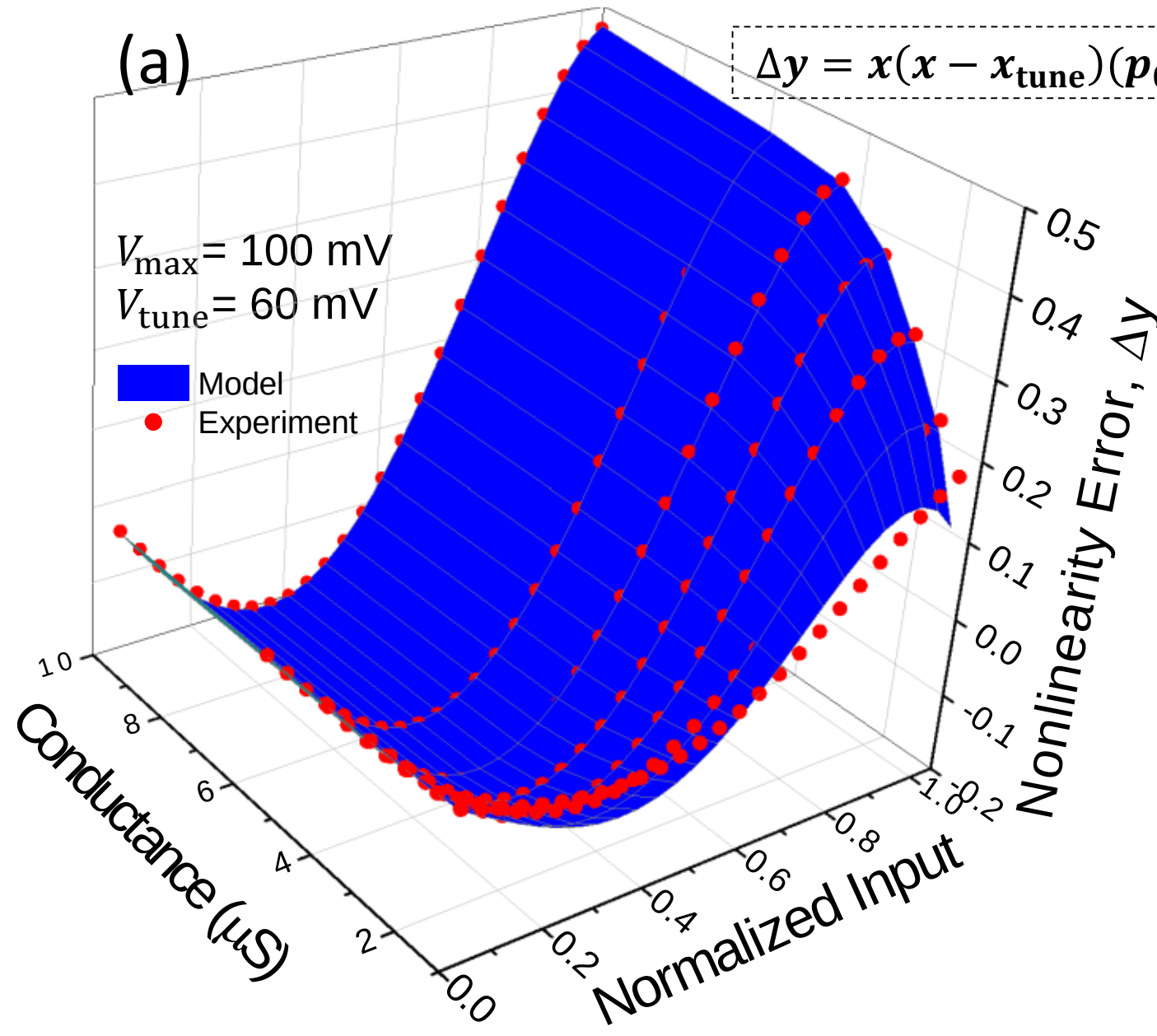

| | | |
|---|---|---|
| Params | $p_{01}$ | 1.368 |
| | $p_{03}$ | -1.662 |
| | $p_{10}$ | 0.7345 |
| | $p_{20}$ | -0.1184 |
| | $p_{30}$ | 0.006132 |
| | $x_{\mathrm{tune}}$ | 0.6076 |
| Goodness of fit | SSE | 0.0054 |
| | R − Square | 0.9990 |
| | Adjusted R − Square | 0.9990 |
| | RMSE | 0.0055 |

$$\Delta y = x(x - x_{\mathrm{tune}})(p_{01}x + p_{03}x^3 + p_{10}w + p_{20}w^2 + p_{30}w^3 + p_{11}xw + p_{22}x^2w^2)$$

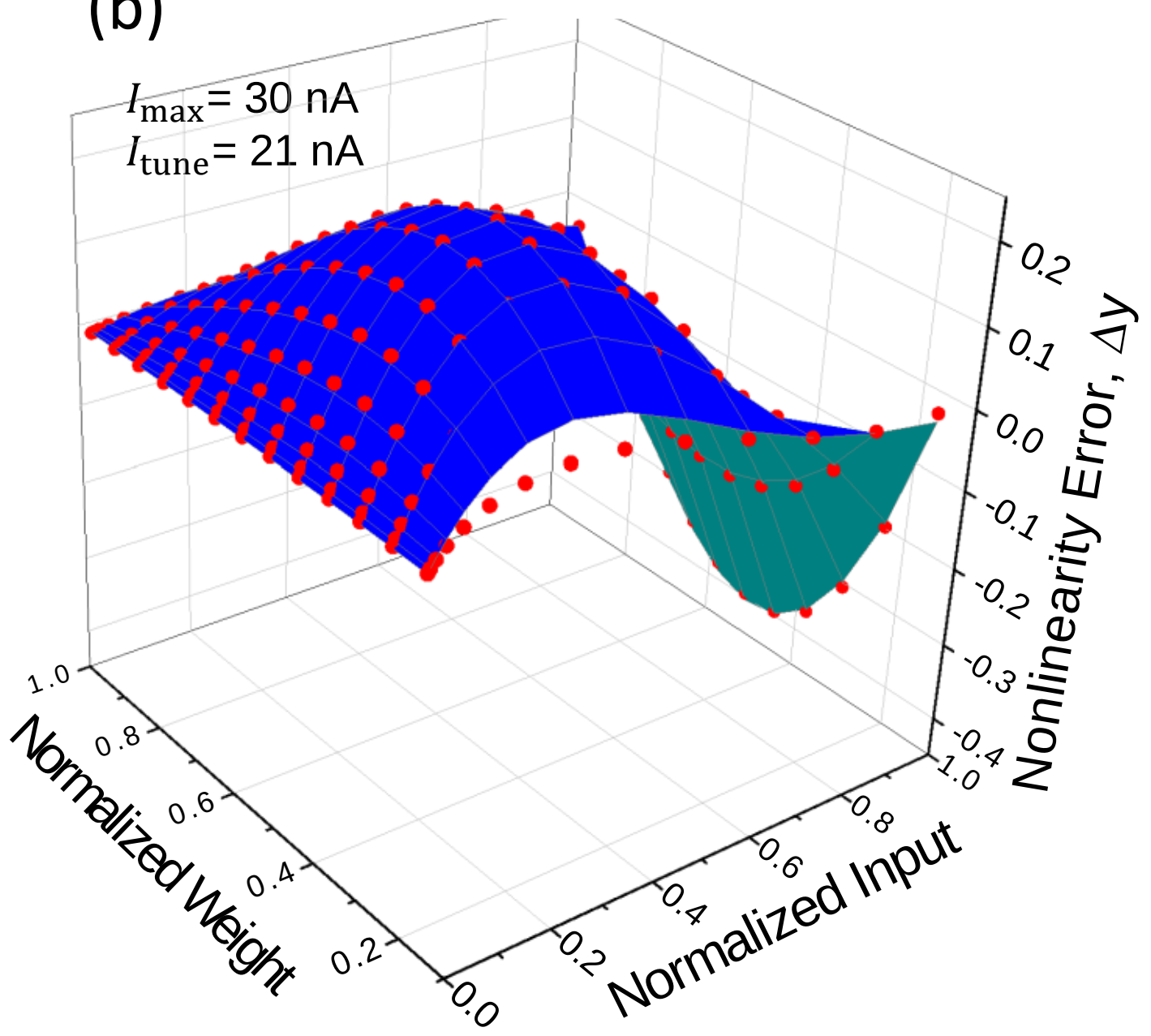

| | | |
|---|---|---|
| Params | $p_{01}$ | 3.109 |
| | $p_{03}$ | -1.222 |
| | $p_{10}$ | -12.71 |
| | $p_{20}$ | 20.41 |
| | $p_{30}$ | -7.816 |
| | $p_{11}$ | -2.733 |
| | $p_{22}$ | 0.9496 |
| | $x_{\mathrm{tune}}$ | 0.7247 |
| Goodness of fit | SSE | 0.002536 |
| | R − Square | 0.9995 |
| | Adjusted R − Square | 0.9995 |
| | RMSE | 0.002665 |

**Supplementary Figure 3.** Modeling nonlinearity in memristor and eFlash synaptic devices. (a) Modeling results for passively integrated memristive devices and (b) redesign eFlash memories. Like the temperature modeling, we opt to use a multi-order polynomial function that perfectly describes devices' average behavior without slowing down massive neuromorphic networks' simulation time. In both panels (a,b), we manually optimize the polynomial functions' shape and use nonlinear least-squares optimization with a trust-region algorithm to obtain the model parameters. Note that in order to ease the network simulation, in this part, we model the nonlinearity error (and not the relative nonlinearity error). The amount of nonlinearity error is a function of both conductances of the device in the tuning biasing condition, maximum applied input signal, and the applied input signals. Hence, to avoid complicating the nonlinearity model and enhance the fitting results, we decouple it from the tuning conditions and maximum applied input signals, i.e., we perform the modeling and find the parameters for each design case once (separately). Here, the results are provided for one case in memristive circuits and one case of eFlash designs. In the former, the error in the synaptic current of a device tuned to $w$ (the conductance of a single device in the differential pair in μS) at the normalized input $x_{\text{tune}}$, when stimulated by $x$ is modeled by $\Delta y = x(x - x_{\text{tune}})(p_{01}x + p_{03}x^3 + p_{10}w + p_{20}w^2 + p_{30}w^3)$. For the latter, since the gate-coupled structure is studied, using both normalized weights and inputs make the modeling easier. Here, when a normalized input $x$ is applied to a synaptic device, tuned to the normalized weight of $w$ at the normalized input $x_{\text{tune}}$, it creates a nonlinearity error that can be obtained by $\Delta y = x(x - x_{\text{tune}})(p_{01}x + p_{03}x^3 + p_{10}w + p_{20}w^2 + p_{30}w^3 + p_{11}xw + p_{22}x^2w^2)$. In both models, $p_{ij}$s are the fitted parameters that are provided in the inset tables. For memristors, the parameters correspond to the case with $V_{\max} = 0.1$ and $V_{\text{tune}} = 0.06$, while for eFlash, $I_{\max} = 30$ nA and $I_{\text{tune}} = 21$ nA, i.e., the devices are tuned at the condition in which the input signals are 0.06 V (for memristors) and 21 nA (for eFlash). The fitting results show excellent goodness of fit across the for both synaptic device candidates.

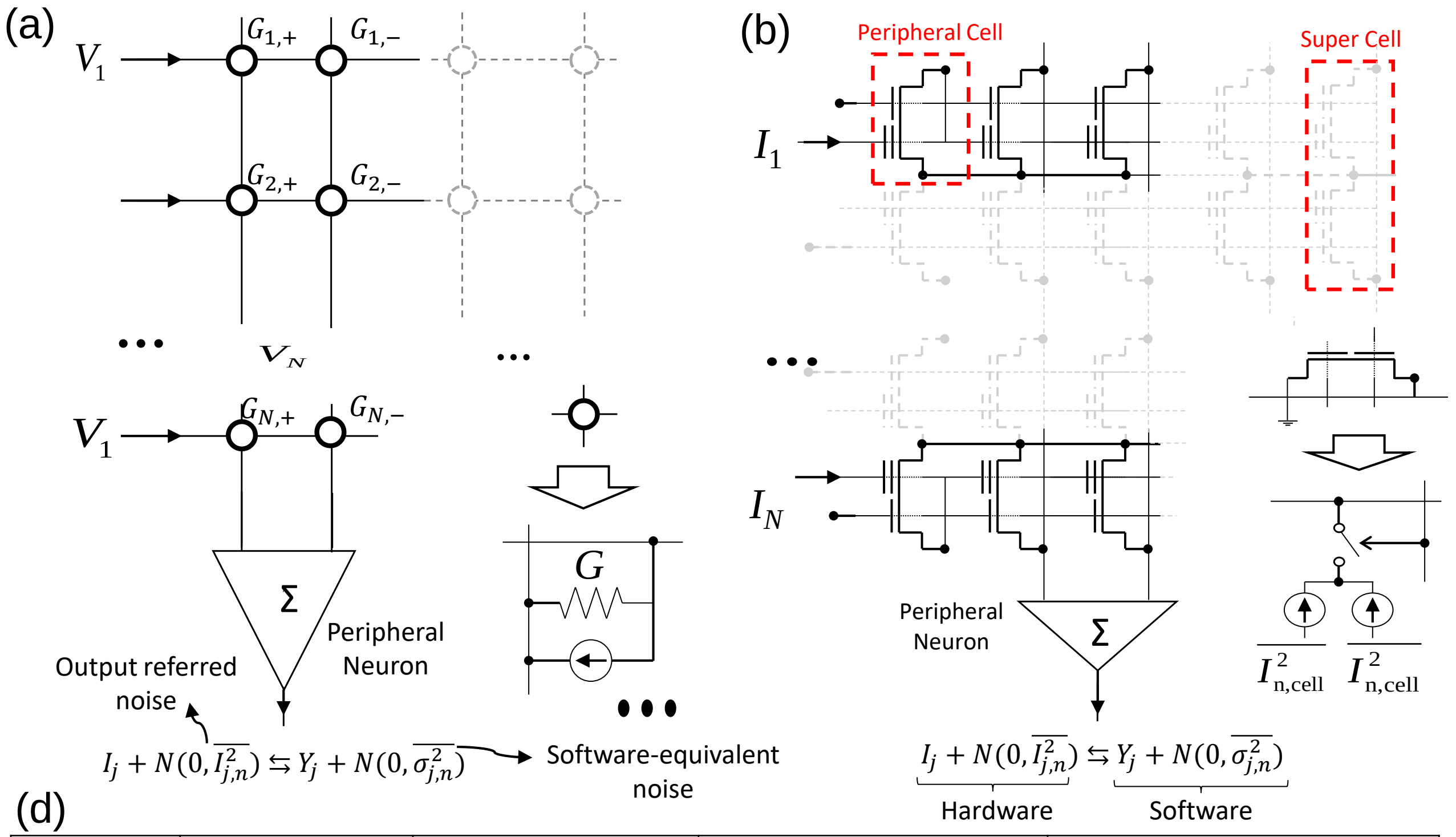

(d)

| | Hardware (Memristors) | Software (Memristors) | Hardware(eFlash) | Software (eFlash) |
|---|---|---|---|---|
| Pre-synaptic Input | $V_i$ | $X_i$ | $I_i$ | $X_i$ |
| Maximum Pre-synaptic Input | $V_{max}$ | $X_{max}$ | $I_{max}$ | $X_{max}$ |
| Synaptic Weight | $G_{ij,+} - G_{ij,-}$ | $W_{ij}$ | $(I_{ij,+} - I_{ij,-})/I_{max}$ | $W_{ij}$ |
| Maximum Abs. Synaptic Weight | $\Delta G_{max}$ | $W_{max}$ | $\Delta I_{max}$ | $W_{max}$ |
| Common-mode Synaptic Weight | $G_{ij,\text{CM}} = G_{ij,+} + G_{ij,-}$ | - | $I_{ij,\text{CM}} = I_{ij,+} + I_{ij,-}$ | – |
| Post-synaptic Output | $I_{ij} = V_i(G_{ij,+} - G_{ij,-})$ | $Y_{ij}=X_iW_{ij}$ | $I_{ij} = \text{I}_i(I_{ij,+} + I_{ij,-})/I_{max}$ | $Y_{ij}=X_iW_{ij}$ |
| Synapse Noise | $\overline{I_{ij,n}^2} = \alpha_M G_{ij,\text{CM}}$ | $\sigma_{ij,n}^2 = \alpha_M G_{ij,\text{CM}}(\frac{X_{max}W_{max}}{V_{max}\Delta G_{max}})^2$ | $\overline{I_{ij,n}^2} = \alpha_F I_i\left(\frac{I_{ij,\text{CM}}}{I_{max}} + (\frac{I_{ij,+}}{I_{max}})^2 + (\frac{I_{ij,-}}{I_{max}})^2\right)$ | $\sigma_{ij,n}^2 = \alpha_F \frac{X_{max}X_i}{I_{max}}\left(\frac{I_{ij,\text{CM}}}{I_{max}} + (\frac{I_{ij,+}}{I_{max}})^2 + (\frac{I_{ij,-}}{I_{max}})^2\right)$ |
| Neuron Pre-activation | $I_j = \sum_{i=1}^{N}(G_{ij,+} - G_{ij,-})V_i$ | $Y_j = \sum_{i=1}^{N} X_iW_{ij}$ | $I_j = \sum_{i=1}^{N} I_i(I_{ij,+} + I_{ij,-})/I_{max}$ | $Y_j = \sum_{i=1}^{N} X_iW_{ij}$ |
| Pre-activation Referred Noise | $\overline{I_{j,n}^2} = \sum_{i=1}^{N} \alpha_M G_{ij,\text{CM}}$ | $\sigma_{j,n}^2 = \alpha_M(\frac{X_{max}W_{max}}{V_{max}\Delta G_{max}})^2 \sum_{i=1}^{N} G_{ij,\text{CM}}$ | $\overline{I_{j,n}^2} = \sum_{i=1}^{N} \alpha_F I_i\left(\frac{I_{ij,\text{CM}}}{I_{max}} + (\frac{I_{ij,+}}{I_{max}})^2 + (\frac{I_{ij,-}}{I_{max}})^2\right)$ | $\sigma_{j,n}^2 = \alpha_F \frac{X_{max}}{I_{max}} \sum_{i=1}^{N} X_i(\frac{I_{ij,\text{CM}}}{I_{max}} + (\frac{I_{ij,+}}{I_{max}})^2 + (\frac{I_{ij,-}}{I_{max}})^2)$ |

**Supplementary Figure 4. Noise analysis in eFlash and 0T1R memristive crossbar circuits.** Noise in (a) memristive crossbar circuits and (b) gate-couple eFlash matrix multipliers (WL, EGs, and SLs are left floating only to make the figure concise). (c) Summary of the noise analysis results in both synaptic devices. Direct circuit simulations in electronic design automation tools are infeasible due to the neuromorphic benchmarks' massive size. This analysis aims to find the software-equivalent noise used to simulate these circuits in the software domain properly. A practical neuromorphic circuit is expected to operate in >100 MHz regimes with dominant white (shot or thermal) noise spectrum, which would be realistic for both eFlash and memristor-based circuits in which arrays are tightly integrated with peripheral circuits. Hence, we consider normally distributed independent noise sources in circuit analysis. Shot noise is expected to dominate the noise performance in weak-inversion biased eFlash memories with intrinsic ballistic transport. The spectral density of the shot noise per unit bandwidth channel is $2\mathrm{q}I$ in eFlash, where $I$ is the synaptic current flowing in the device, and q is the electron charge. Shot noise is negligible in memristors, and thermal noise is dominant, owing to the diffusive electron transport and relatively small applied voltage. The thermal noise can be represented in the circuit by a parallel current source with the spectral density $4\mathrm{K}TG$ where $K$= 1.38 × $10^{-23}$ J/K is the Boltzmann constant, $T$ is the temperature in Kelvin, and $G$ is the device conductance. It is noteworthy some devices also exhibit a considerable random telegraph noise, which can be added to the thermal noise power. Given the fact that noise sources are uncorrelated and independent, we find the preactivation referred current noise of a neuron in an $N$-input circuit by adding the noise power of each synaptic device that effectively turns out to $\overline{I_{ij,\mathrm{n}}^2} = \alpha_\mathrm{M} G_{ij,\mathrm{CM}}$, where $\alpha_\mathrm{M} = 4\mathrm{K}TB_0$ in which $B_0$ is an equivalent noise bandwidth, $G_{ij,\mathrm{CM}} = G_{ij,+} + G_{ij,-}$ is the common-mode conductance of a differential pair, and the rest of the parameters have their usual meaning. Here, we neglect the input-referred noise from the peripheral circuits to primarily focus on synapses. Given the white dominant noise spectrum, the sensed preactivation signal by the neuron is approximated by $I_j + \mathrm{N}(0, \overline{I_{j,\mathrm{n}}^2})$ in hardware which corresponds to $Y_j + \mathrm{N}(0, \overline{\sigma_{j,\mathrm{n}}^2})$ in the software domain. To find the software-equivalent noise ($\sigma_{j,\mathrm{n}}^2$), we note a linear mapping of input voltages in the circuit to input signals in software, i.e., $X_i/X_\mathrm{max} = V_i/V_\mathrm{max}$. Besides, it is critical to ensure that the signal-to-noise ratio of a synaptic device remain equal in both software and hardware, i.e., ${I_{ij}}^2/I_{ij,\mathrm{n}}^2 = {Y_{ij}}^2/\sigma_{ij,\mathrm{n}}^2$. Using these equations, we find the preactivation noise variance $\sigma_{j,\mathrm{n}}^2 = \alpha_\mathrm{M}(X_\mathrm{max}W_\mathrm{max}/V_\mathrm{max}\Delta G_\mathrm{max})^2 \sum_{i=1}^{N} G_{ij,CM}$. A similar analysis is performed for the eFlash-based circuits that lead to $\sigma_{j,\mathrm{n}}^2 = \alpha_F(X_\mathrm{max}/I_\mathrm{max}) \sum_{i=1}^{N} X_\mathrm{i}(I_{ij,\mathrm{CM}}/I_\mathrm{max} + (I_{ij,\mathrm{CM}}/I_\mathrm{max})^2)$ in which $\alpha_\mathrm{F} = 2\mathrm{q}B_0$. We define an energy scaler variable $\rho$ and include it as a multiplicative factor in the software equivalent noise equations for simulation purposes. $\rho$ =1 corresponds to nominal operating conditions, i.e., 100 MHz bandwidth, $V_\mathrm{max}$=0.1, $\Delta I_\mathrm{max} = 9$ µA in our memristors and $\Delta I_\mathrm{max}$=30 nA in eFlash memories. The benefit of using this unified scaling factor lies in two folds. On one hand, we can represent the simulation trends without delving into the details of changing bandwidth, power consumption, or dynamic range. Reducing $\rho$ simply implies less energy is spent per device (either by slowing down or reducing the dynamic range). On the other hand, the energy scaling factor simplifies how downscaling memristor technology changes fidelity trends. For example, $\rho$=0.25 corresponds to 4× slower operation or using 2× higher (voltage or conductance) dynamic range, or ~4× finer technology node (which is expected to reduce the common-mode conductance range roughly by a factor of 4). Several informative observations from the software-equivalent noise equations are noteworthy. First, the noise variance in passive memristor circuits is independent of the input signal due to the fact that memristive devices in a neuron always contribute to the corresponding preactivation noise regardless of the applied voltage/input. On the contrary, the noise in eFlash memory circuits is input dependent, and synapses that conduct zero current do not contribute any noise to the output. For both cases, the noise depends on the mapping: in mapping 1, the common-mode synaptic weight depends on the weight (e.g., $G_{ij,CM} = 2G_{ij,\mathrm{min}} + \Delta G|W_{ij}|/W_\mathrm{max}$), but in mapping 2, it is constant (e.g., $G_{ij,CM} = 2G_\mathrm{b}$). Finally, two knobs change the noise power or variance: circuit parameters such as power consumption and network parameters such as maximum weight and input magnitudes, which we exploit in our proposed approaches 1 and 2 to improve the performance.

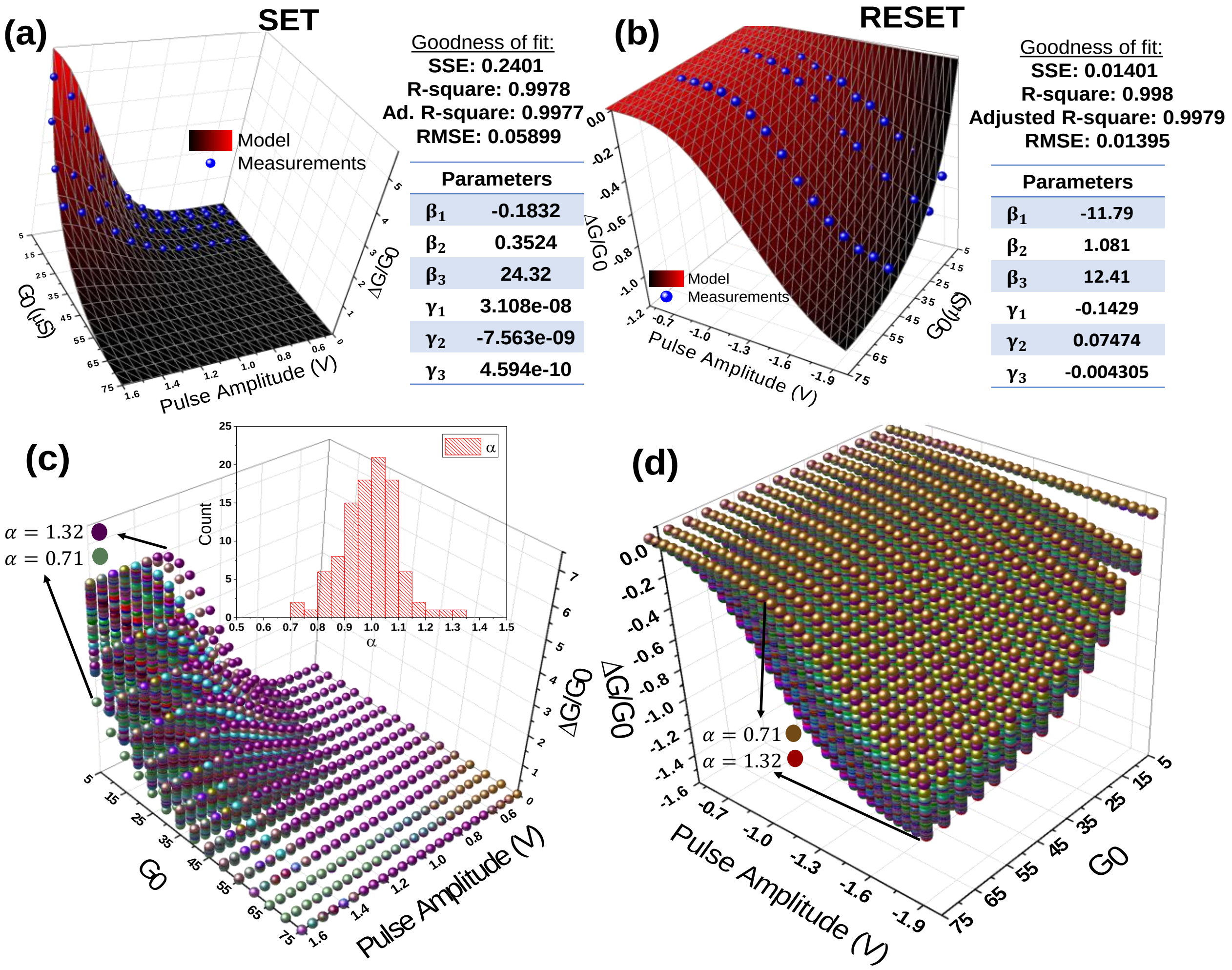

**Supplementary Figure 5. Dynamic model and tuning precision study.** The tuning precision in analog memories with selectors, e.g., eFlash, is typically determined by how precisely we can adjust the state of individual devices, which is a function of their switching characteristics, retention, tuning algorithm, etc. However, in passive memories, the half-select problem changes the dynamics of tuning and create a large tail of devices with tuning errors higher than the target error, in particular when the variations are high, or the crossbar size is large. To properly simulate this issue, we develop a dynamic model that predicts our memristors' dynamic behavior and emulates the tuning and ex-situ weight transfer processes. The model predicts the change in a device's conductance as a function of its initial state and the pulse amplitude (the duration is fixed at 2 ms to simplify the model). We use a trust-region algorithm for nonlinear least-squares to fit experimental data points from 500 devices to $\Delta G/G \approx \exp[\beta_1/(1+\beta_2(\alpha V)^2)]\sinh[\beta_3\alpha V/(1+\beta_2(\alpha V)^2)]\left(\gamma_1+\gamma_2\sqrt{G}+\gamma_3 G\right)$, where $G$ is the initial conductance, $\beta_i$ and $\gamma_i$ are fitting parameters, and α is a device-unique multiplicative factor that models the variations in the switching thresholds. Panels (a) and (b) show the modeling results for the average set and reset operations, respectively. The model parameters closely reproduce the measurement results. The inset tables show the corresponding goodness of the fit and model parameters. Panels (c-d) show the set/reset characteristics for 100 devices with 10% normalized variations. The inset shows the corresponding distribution of α.

In order to emulate the tuning process of a crossbar (ex-situ mapping of weights to the conductance of memristors, while considering the tuning error), we randomly initialize the conductances of devices using a Gaussian distribution with an average of 36.25 µS (midrange conductance) and a standard deviation of 9 µS. This is to assume prior to beginning the tuning process; the devices are randomly distributed in the allowable conductance range. Then, we sequentially tune each device's conductance using the write-verify algorithm and the developed dynamic model. This is performed by emulating the exact procedure employed in the experiments when tuning the actual devices: The devices within any crossbar block are tuned in raster order. More importantly, to increase the tuning speed, we progressively increase the pulse amplitude (set/reset) starting from 0.5 V with 10 mV steps to the device's switching voltage. This is to avoid overstressing the device. The tuning direction (setting or resetting) is alternated whenever we pass the target conductance. To avoid overstressing the memristors, creating too much disturbance, and reducing the tuning time, we limit this to 5 rounds. The algorithm is aborted (and restarted with the next device) whenever it reaches the desired tuning accuracy or the maximum permitted pulse per device. Note that after every single pulse, the half-select disturbance is applied by updating the state of devices sharing either top/bottom electrode with the V/2 rule. Finally, the entire kernel is tuned in 10 rounds to minimize the disturbances as much as possible.

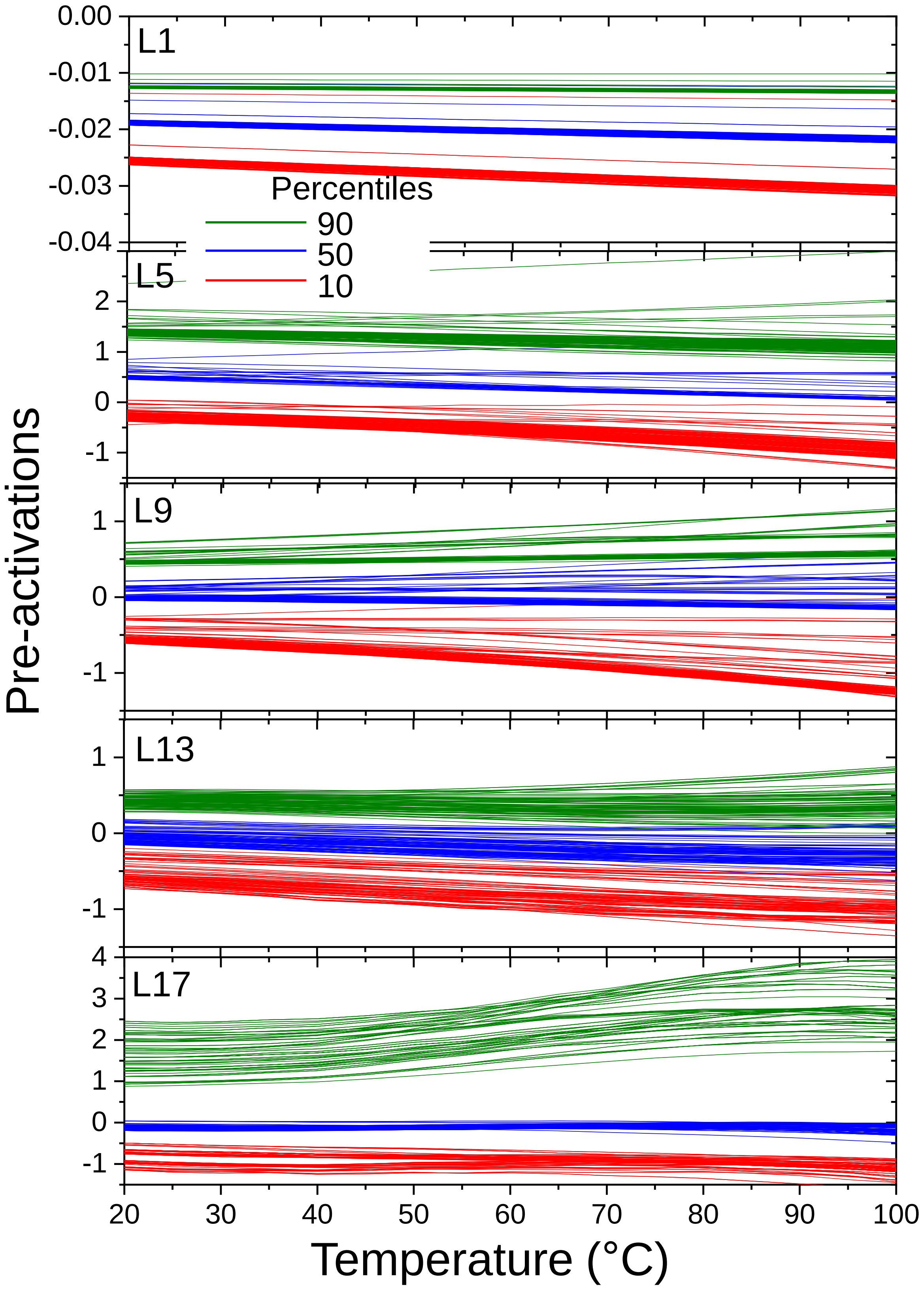

**Supplementary Figure 6. Preactivation statistics versus temperature.** An almost monotonic shift in 10, 50, and 90 percentiles of preactivation statistics in 100 randomly selected neurons in various ResNet-18 layers. The statistics are obtained by processing 100 batches, and the temperature model of the RRAM, mapping 1 is used.

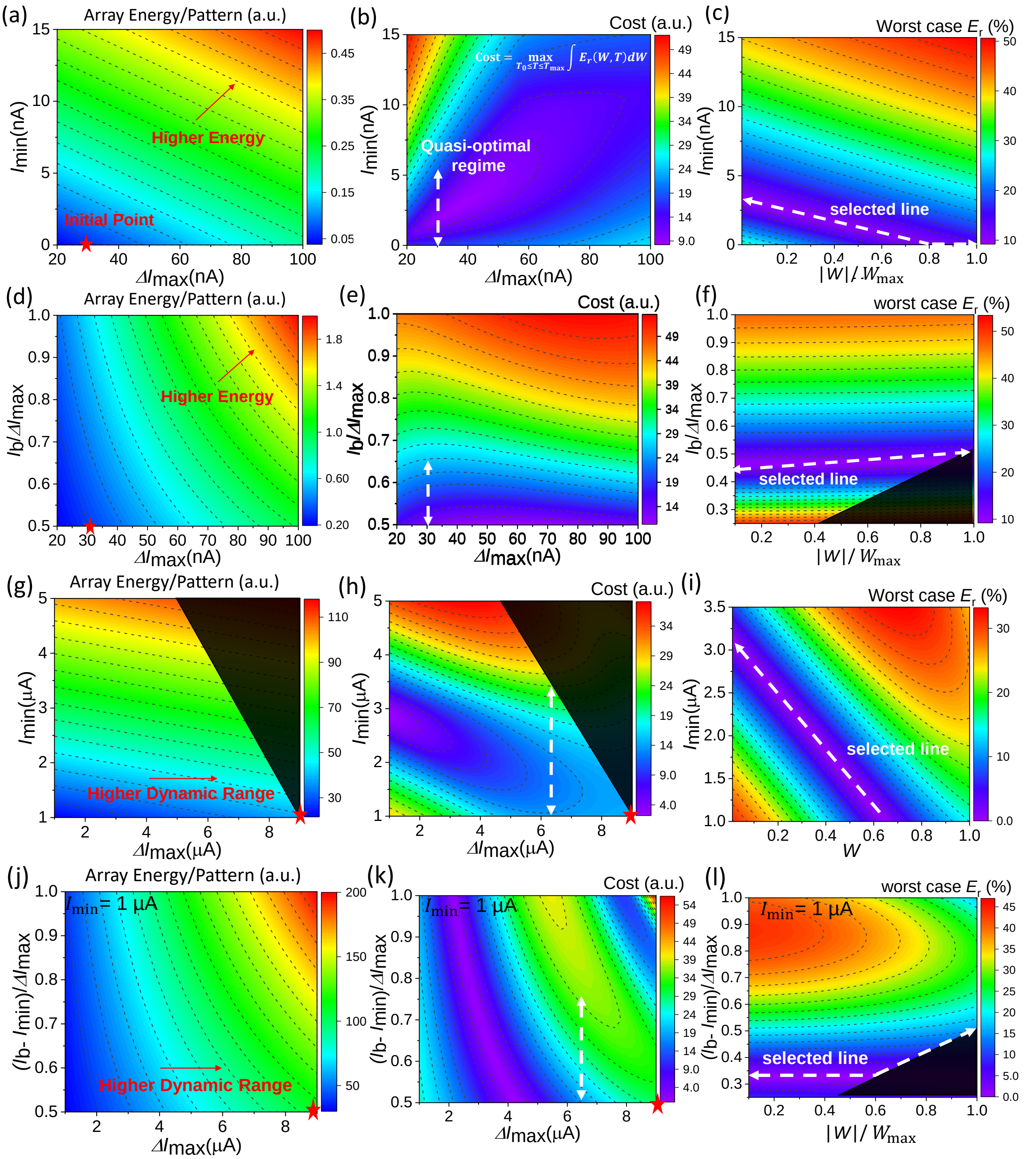

**Supplementary Figure 7. State optimization for temperature sensitivity.** Panels (a-c), (d-f), (g-i), and (j-l) show the state optimization simulation results corresponding to mapping 1 of eFlash, mapping 2 of eFlash, mapping 1 of memristors, and mapping 2 of memristors, respectively. The shaded areas denote out-of-range regimes. The panels in the first column show the normalized energy consumption in synaptic arrays for a network of 10M normally distributed weights versus the dynamic range ($\Delta I_{max}$). The choice of the initial design point ($\Delta I_{max}$, $I_{min}$, and $I_b$) is often in the direction of minimizing the energy consumption in eFlash circuits and maximizing the dynamic range in 0T1R memristive systems, regardless of the mapping type (the red star in the first column panels shows the initial design point). However, the optimum sensitivity concerning temperature variations is not necessarily this design point. Since we intend to apply a secondary cost-free technique to further compensate for temperature variations, our goal in this step is to trade energy or dynamic range to improve the accuracy and find a quasi-optimum operating point that is less sensitive to temperature variations. To find a quasi-optimum design point, we define the cost function $C = \max_{T_0 \le T \le T_{\max}} \int_{\in^+}^{1} |E_r(W,T)| dW$, which represents the worst sum of relative errors among all (normalized) weights across all temperatures. To numerically compute the cost function, we use $T_0$=25, $T_{\max}$=100, and $\in^+$=0.01. The minima of the cost function give the optimum design point averaged over weights. The panels in the second column show how $C$ changes across the design space. Based on the heatmap of *C*, we can select $\Delta I_{\max}$ that is close to the minima without overspending on energy (in eFlash) or dynamic range (in memristors). By definition, $\Delta I_{\max}$ is weight independent; however, we may optimize the other design parameter ($I_{\min}$ or $I_{\text{b}}$) at the cost of slight power increase or dynamic range reduction. Unlike previous works that choose a fixed minimum current or bias current for all weights, we find a more optimum weight-dependent choice of minimum synaptic current (mapping 1) and bias current (mapping 2) by using third column panels that show the heatmap of the worst-case relative error across all temperatures versus normalized weight. Panel (b) shows the cost function for eFlash mapping 1. A white dashed line ($\Delta I_{\max}$=30 nA) indicates a quasi-optimal regime that features low energy and is close to the minima of *C*. The error is further optimized by finding an optimum weight-dependent $I_{\min}$. Panel (c) shows that the worst-case error is minimum when $I_{\min}(\text{nA}) = \max(0, 3 - 3.75(|w|/w_{\max}))$. For mapping 2 (second row panels), we observe that the cost function and energy are both minimized when the minimum bias current is used, i.e., $I_{\text{b}} = \Delta I_{\max}/2$. To minimize power, $\Delta I_{\max} = 30$ nA (the same as mapping 1) is used, and the optimum bias current for a given weight is obtained by $I_{\text{b}}(\text{nA}) = 2.35(|w|/w_{\max}) + 12.65$. The same procedure is used for the memristors, and similar results are obtained. Operating eFlash in deep weak inversion enables low power operation and high dynamic range. Hence, trading a slight increase in energy consumption for improved reliability makes a lot of sense. Unlike eFlash devices (at least in the present technology), metal-oxide memristors are more power-hungry and have a limited dynamic range, limiting the options for finding the quasi-optimized state. In mapping 1 of memristors (panel h), we observe that the minimum cost is obtained in a region that has a very low dynamic range (which is impractical to tune the weights and realistically map the weights to it). Instead of using a low dynamic range, we choose a practically viable dynamic range (6.5 µA) and reserve 3.5 µA for finding an optimum weight-dependent $I_{\min}$. Panel (i) shows that $I_{\min}(\mu\text{A}) = \max(0, 3.1 - 3.23(|w|/w_{\max}))$ is the weight-dependent quasi-optimal equation for our devices for the 6.5 µA dynamic range. Similarly, 6.5 µA dynamic range is selected for mapping 2, and the optimum bias current per weight is obtained by $I_{\text{b}}(\mu\text{A}) = \max(3.1, 1.48 + 2.76(|w|/w_{\max}))$.

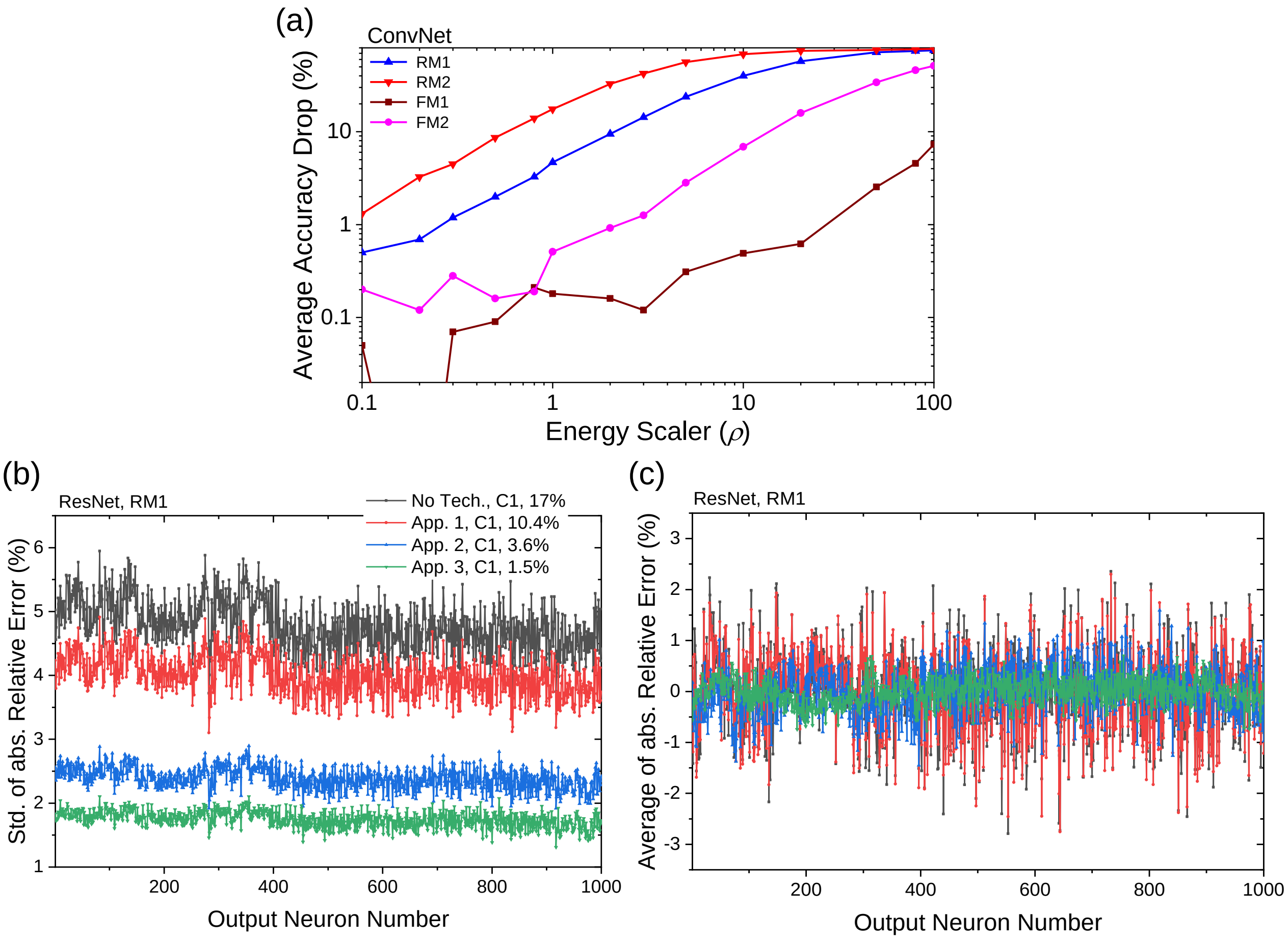

**Supplementary Figure 8. Extended simulation data on noise study.** (a) The average accuracy drop in ConvNet versus the global energy scaler (the same energy scaler is used in all layers) for RM1 (ReRAM, Map1), RM2 (ReRAM, Map2), FM1 (eFlash, Map1), and FM2 (eFlash, Map2). (b) The standard deviation and (c) the average of the absolute relative error in the outputs of ResNet-18 (RM1) model for a case with 17% initial accuracy drop, and after the proposed approaches are applied incrementally. The statistics are obtained by testing ~500k samples.

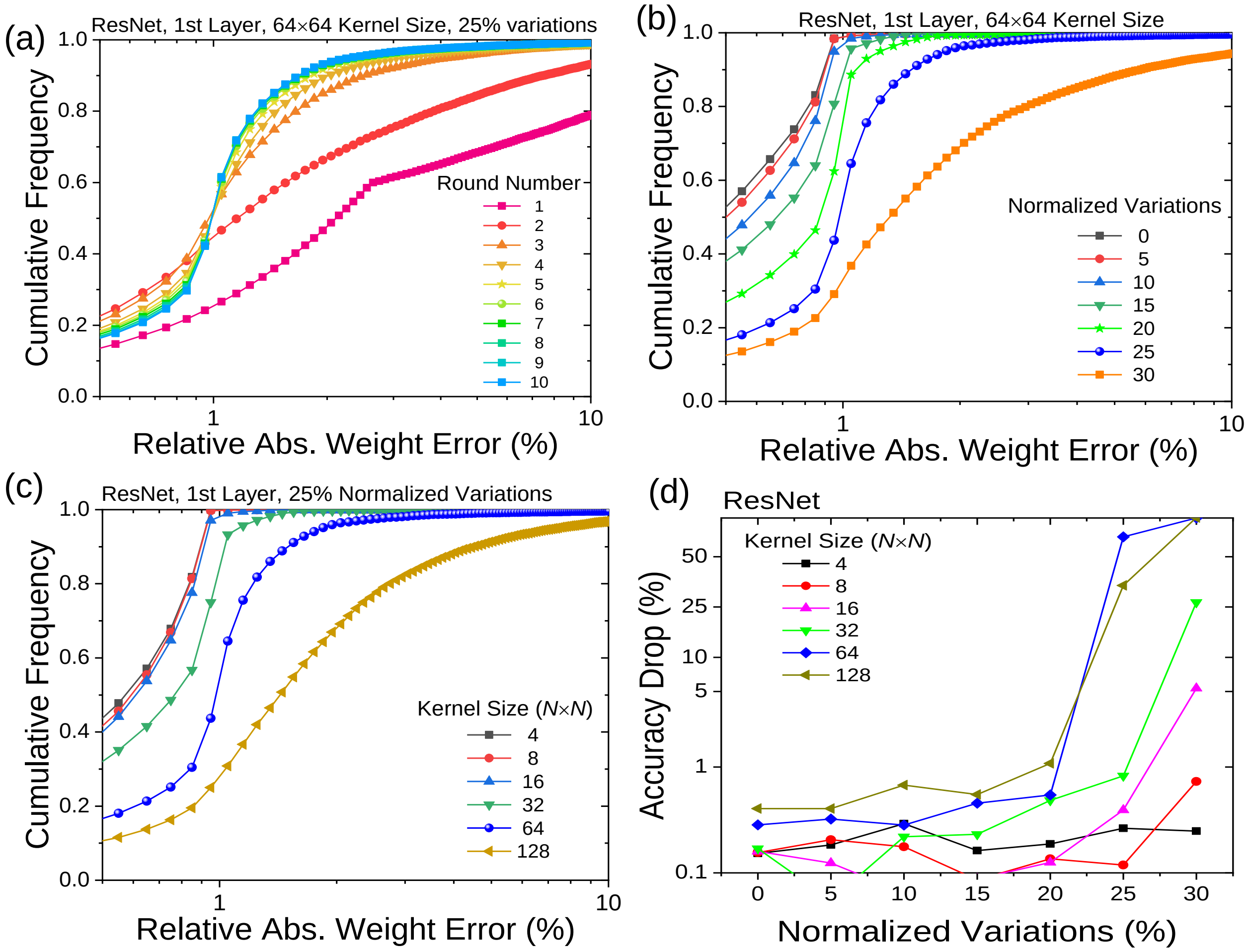

**Supplementary Figure 9. Extended data on half-select disturbance simulations.** (a) The normalized cumulative distribution of tuning error in consecutive rounds of running the tuning algorithm. The data corresponds to the first layer of ResNet-18 (~2×9.5k devices), and the case study of employing 64×64 crossbars and a technology with 25% normalized switching threshold variations. (b) The normalized cumulative distribution of relative tuning error for various normalized variations. The data corresponds to the first layer of ResNet-18 and the case study of employing 64×64 crossbars. (c) The normalized cumulative distribution of relative tuning error for various kernel (crossbar) sizes. The data corresponds to the first layer of ResNet-18, and the case study of employing a technology with 25% normalized switching threshold variations. (d) The accuracy drop in ResNet-18 for various kernel sizes and switching threshold variations. In all simulations, we consider a 1% target tuning error verified by experiments in Fig. 2 of the main text.

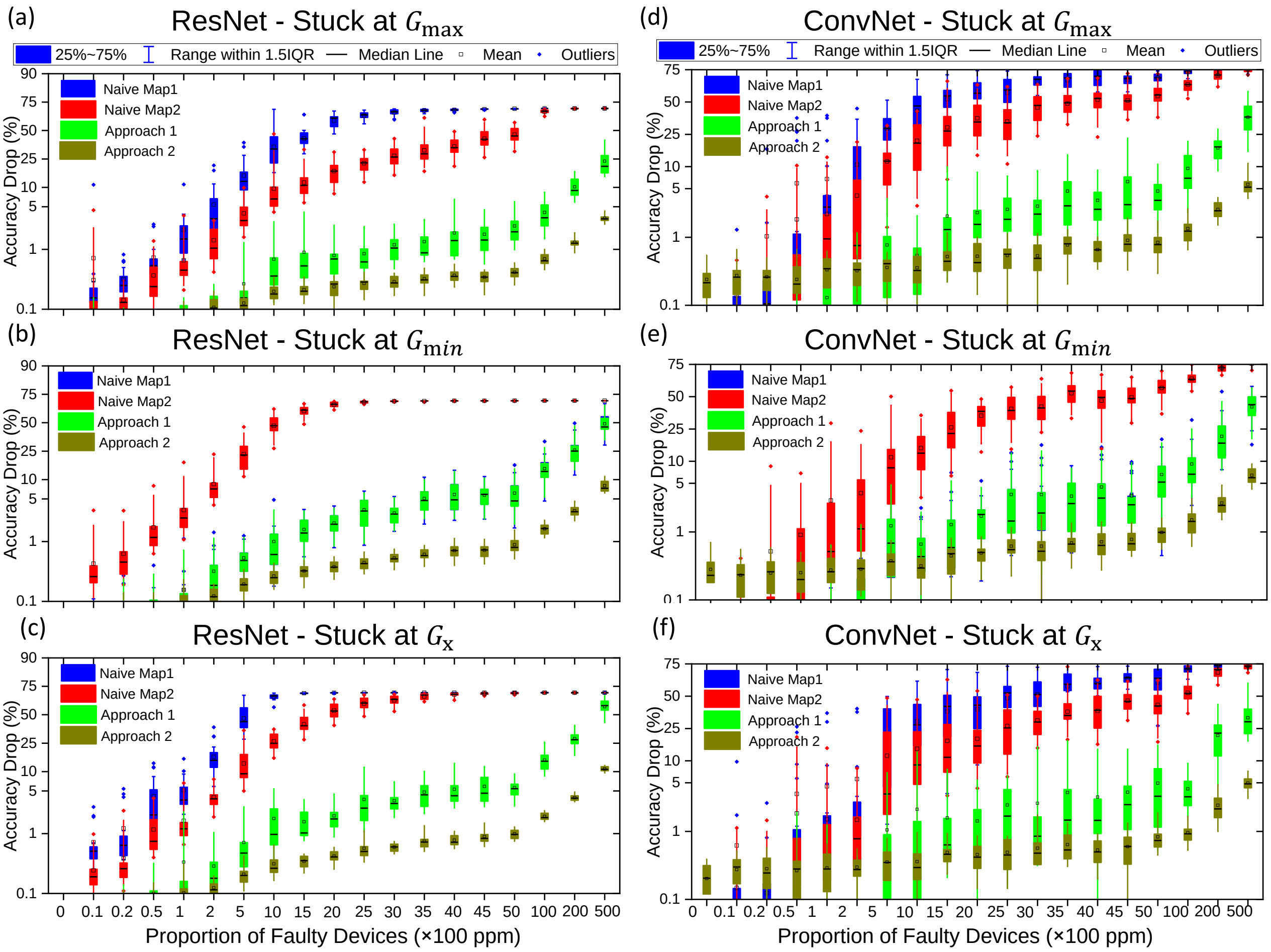

**Supplementary Figure 10. Extended yield Analysis simulation results.** Fault-tolerance analysis in (a-c) ResNet-18 and (d-f) ConvNet. In panels (a,d), the defective devices are stuck at high conductance ($G_{\max}$). In panels (b,e), the defective devices are stuck at low conductance ($G_{\min}$), and in panels (c,f), the conductance of faulty devices are uniformly distributed in the considered conductance range ($G_{\min} < G_{\mathrm{x}} < G_{\max}$). For every point, the statistics are obtained over 20 runs. The locations of defective devices in all simulation points are randomly selected. The results of approach 1 (and approach 2) are the same for both mappings because of using the same compensation scheme (see the method section in the main text) in pair-wise adjustment independent of the original mapping scheme.

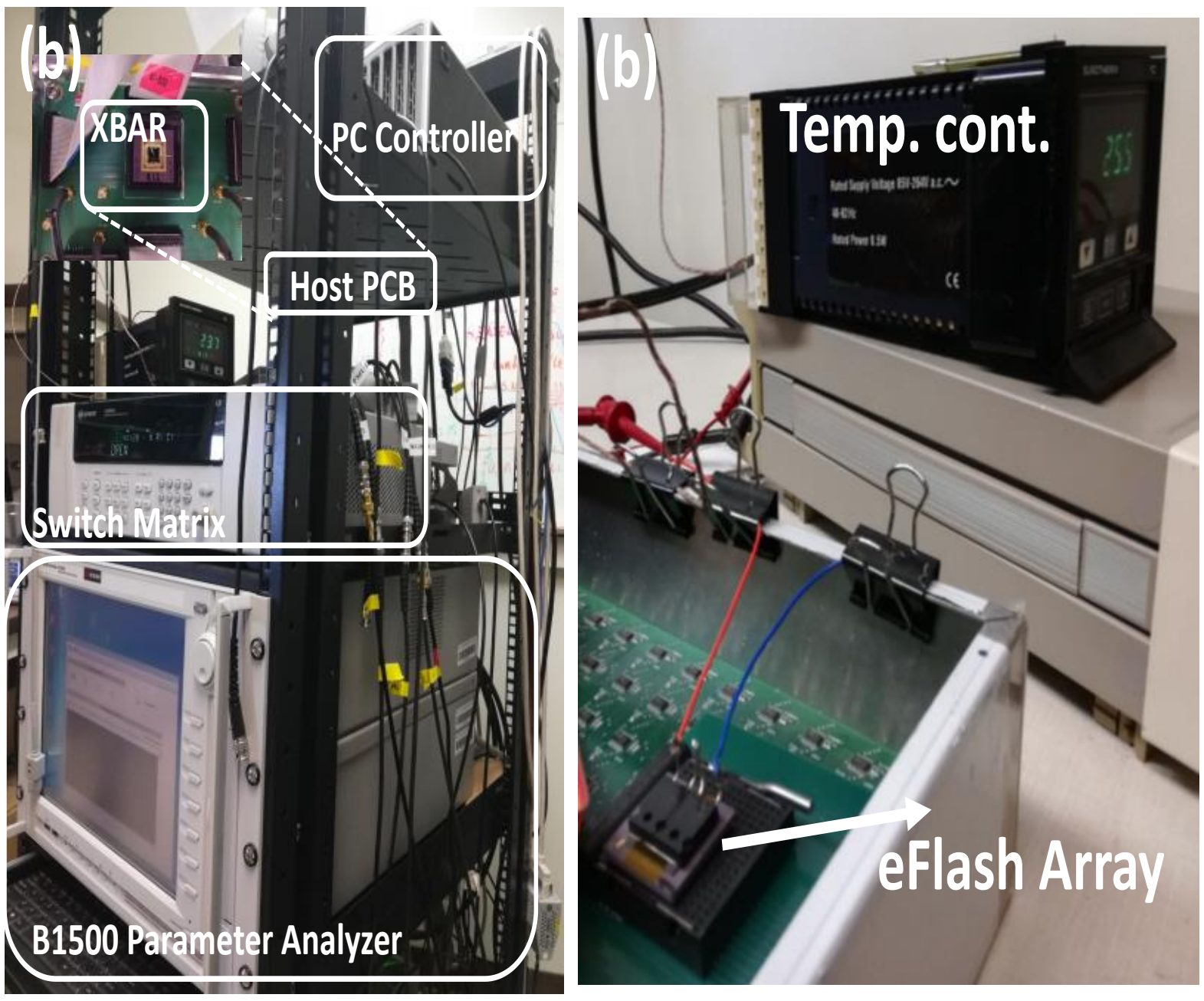

**Supplementary Figure 11. Experimental setup.** (a) The memristor characterization setup includes the packaged crossbar mounted on a custom-designed host printed circuit board, Agilent switch matrix, parameter analyzer, and personal computer, which is controlled by the user. (b) The eFlash characterization setup, including a custom-designed switch matrix, eFlash chip, and temperature controller.

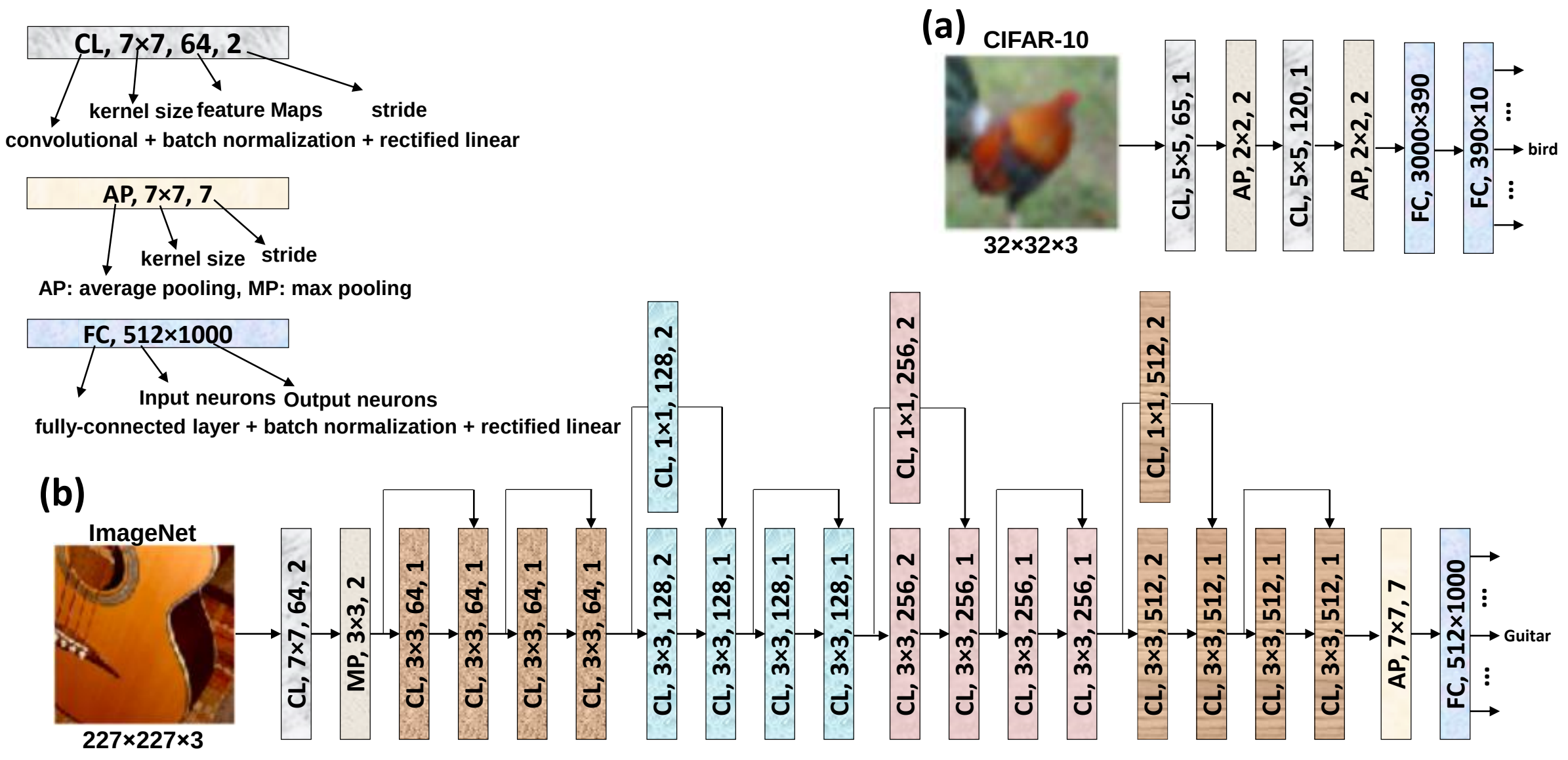

**Supplementary Figure 12. The neuromorphic benchmarks.** (a) A 6-layer convolutional neural network trained on the CIFAR-10 dataset. (b) The ResNet-18 model trained on the ImageNet dataset.

| Ref | Device Stack | Exp. or sim | Benchmarks | Studied Imperfections* | Method |
|---|---|---|---|---|---|
| [1] | ReRAM | Hybrid | MLP (MNIST) | N | Impact investigation, No mitigation proposed. |
| [2] | ReRAM | Sim | VMM | IR | Adding external resistors in peripheries for average error compensation |
| [3] | ReRAM | Sim | VMM | NL, IR | Impact investigation, No mitigation proposed. |
| [4] | ReRAM | Hybrid | MLP (MNIST) | N, IR | Averaging the outputs of multiple networks using a committee machine |
| [5] | ReRAM | Hybrid | MLP (MNIST) | N | Impact investigation, No mitigation proposed. |
| [6] | ReRAM | Sim | MLP (MNIST) | FT | Chip-specific fault-tolerant mapping |
| [7] | ReRAM | Hybrid | MLP (MNIST) | NL, FT, PNL, PE | Impact investigation, No mitigation proposed. |
| [8] | ReRAM | Hybrid | ResNet (ImageNet, (CIFAR-10) | N, R, PE | Partial On-chip calibration and hardware-aware training |
| [9] | ReRAM | sim | MLP (MNIST) CNN(CIFAR10) | FT, PE | Partial On-chip calibration |
| [10] | ReRAM | sim | SLP (MNIST) | IR, PE | Chip-specific calibration |
| [11] | ReRAM | sim | SLP (MNIST) | FT, PE | Partial On-chip calibration |
| [12] | eFlash | Hybrid | CNN(CIFAR10) | N | hardware-aware training |
| [13] | ReRAM | Hybrid | VMM MLP(MNIST) | IR, NL | Algorithmic weight conversion |
| [14] | ReRAM | sim | MLP(MNIST) | FT | Retraining and weight remapping |
| [15] | ReRAM | sim | Lenet(MNIST) ResNet (CIFAR) | FT | Impact investigation and shows Drop connect regularization helps with FT |
| [16] | ReRAM | Hybrid | MLP (MNIST) | FT, PE | Chip-specific calibration, Temperature Compensation in neurons |
| [17] | ReRAM | Hybrid | MLP (MNIST) | IR, NL | Bootstrapping and tuning optimization |
| This Work | eFlash, ReRAM | Hybrid | ResNet (ImageNet) Conv. DNN (CIFAR-10) | T, N, R, NL, PE, FT | A holistic fully ex-situ approach that includes novel modifications in the training, tuning algorithm, state optimization, and circuit design |

* T: Temperature, N: Noise, PE: programming error, in 0T1R, FT: Fault-tolerance, NL: Nonlinearity, R: Retention, IR: IR drop, PNL: programming nonlinearity.

**Supplementary Figure 13. A Summary of previous works.** The table compares this paper with previous works, which have focused on the study of the impact of imperfections in mixed-signal neuromorphic circuits or vector-by-matrix multipliers.